\RequirePackage{tkz-euclide}
\documentclass[pdflatex,iicol]{sn-jnl}

\usepackage{lscape,lipsum}
\usepackage{threeparttable}
\usepackage{longtable}
\usepackage{multirow}
\usepackage{makecell}
\usepackage{etoolbox}
\usepackage{pgfplots} 
\usepackage{tkz-euclide}
\usepackage{graphicx, adjustbox}
\usepackage{subfig}
\usepackage{hyperref}
\usepackage{mathtools}
\usepackage{multirow}
\usepackage{array}
\usepackage{booktabs}
\usepackage[usestackEOL]{stackengine}
\usepackage{xfrac}
\usepackage{threeparttable}
\usepackage{pdflscape}
\usepackage{amsmath} 

\usetikzlibrary{shapes.geometric}
\usetikzlibrary{3d}

\usepackage{program}

\usepackage{geometry}
\graphicspath{{./Images/}}

\usepackage{xcolor}
\definecolor{myred}{RGB}{226,13,118}
\definecolor{myblue}{RGB}{0,112,196}
\definecolor{mynaiveblue}{RGB}{117,232,255}
\begin{document}

\title[Article Title]{Deep Learning in Spatially Resolved Transcriptomics: A Comprehensive Technical View}


\author[1]{\fnm{Roxana} \sur{Zahedi Nasab}}

\author[1]{\fnm{Mohammad Reza} \sur{Eftekhariyan Ghamsari}}

\author[2]{\fnm{Ahmadreza} \sur{Argha}}
\author[3]{\fnm{Callum} \sur{Macphillamy}}
\author[4]{\fnm{Amin} \sur{ Beheshti}}
\author[5]{\fnm{Roohallah} \sur{ Alizadehsani}}
\author[2]{\fnm{Nigel} \sur{ H. Lovell}}
\author[6,7]{\fnm{Mohammad} \sur{Lotfollahi}}
\author*[1,8,9] {\fnm{Hamid} \sur{Alinejad-Rokny}}\email{h.alinejad@unsw.edu.au}

\affil*[1]{BioMedical Machine Learning Lab (BML), The Graduate School of Biomedical Engineering, UNSW Sydney, Sydney, NSW, 2052, Australia}

\affil[2]{The Graduate School of Biomedical Engineering, UNSW Sydney, Sydney, NSW, 2052, Australia}

\affil[3]{School of Animal and Veterinary Sciences, University of Adelaide, Roseworthy, SA, 5371, Australia}

\affil[4]{School of Computing, Macquarie University, Sydney, 2109, Australia}

\affil[5]{Institute for Intelligent Systems Research and Innovation (IISRI), Deakin University, Waurn Ponds, Melbourne, VIC, 3216, Australia}
\affil[6]{Computational Health Center, Helmholtz Munich, Germany}
\affil[7]{Wellcome Sanger Institute, Cambridge, UK}
\affil[8]{UNSW Data Science Hub, The University of New South Wales (UNSW Sydney), Sydney, NSW, 2052, Australia}
\affil[9]{Health Data Analytics Program, AI-enabled Processes (AIP) Research Centre, Macquarie University, Sydney, 2109, Australia}
	
	\abstract{
		Spatially resolved transcriptomics (SRT) has evolved rapidly through various technologies, enabling scien-
		tists to investigate both morphological contexts and gene expression profiling at single-cell resolution in
		parallel. SRT data are complex and multi-modal, comprising gene expression matrices, spatial information,
		and often high-resolution histology images. Because of this complexity and multi-modality, sophisticated
		computational algorithms are required to accurately analyze SRT data. Most efforts in this domain have
		been made to utilize conventional machine learning and statistical approaches, exhibiting sub-optimal results
		due to the complicated nature of SRT datasets. To address these shortcomings, researchers have recently
		employed deep learning algorithms including various state-of-the-art methods mainly in spatial clustering,
		spatially variable gene identification, and alignment. While great progress has been made in developing deep learning-based models for SRT data analysis, further improvement is still needed to create more biologically aware models that consider aspects such as phylogeny-aware clustering or the analysis of small histology image patches. Additionally, strategies for batch effect removal, normalization, and handling overdispersion and zero inflation patterns of gene expression are still needed in the analysis of SRT data using deep learning methods. In this paper, we provide a comprehensive overview of these deep learning methods, including their strengths and limitations. We also highlight new frontiers, current challenges, limitations, and open questions in this field. Also, we provide a comprehensive list of all available SRT databases that can be used as an extensive resource for future studies.}
	
	
	
	
	\maketitle

		\section{Introduction}
		\label{sec1}
\begin{figure*}
			\centering
			\begin{adjustbox}{max width=\textwidth}
				\begin{tikzpicture}
					\node at (12.5,5) {\includegraphics[scale=1.2]{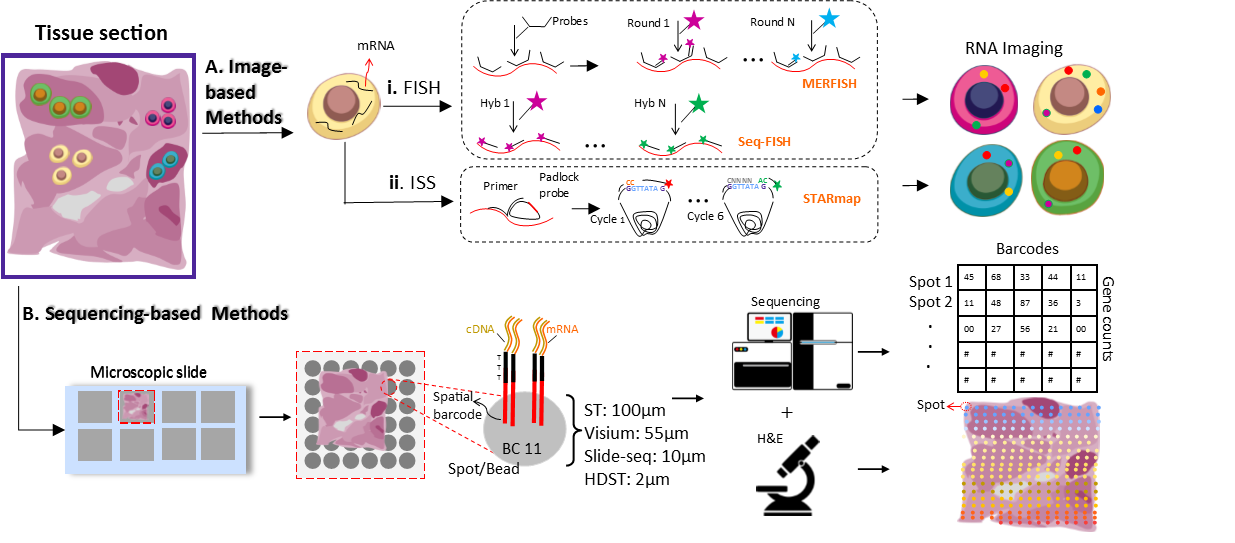}};
				\end{tikzpicture}
			\end{adjustbox}
			\caption{Schematic overview of two SRT approaches. \textbf{A) Image-based methods}. These methods contain two categories. \textit{i. fluorescent in situ hybridization (FISH) approach:} In this category, the probes are labelled with a set of fluorophores that are then individually hybridized to predefined RNA targets to visualize gene expression in fixed tissue. This approach has been further developed to smFISH by utilizing multiple shorter probes, which provide quantitative measurements of transcripts. In 2014, a further development of smFISH involved using sequential hybridizations (seqFISH). To avoid the extensive time of seqFISH, a multiplexed error-robust FISH (MERFISH) was proposed in which N rounds of fluorescence readouts can encode each mRNA by a binary code, and the targeted transcript can be distinguished by decoding. \textit{ii. In situ sequencing methods (ISS):} In this approach, RNA sequencing is performed directly on the RNA content while it remains in its tissue context, mainly using padlock probes to target the genes. For instance, STARmap is an ISS method; it uses six cycles of barcoded padlock probes and adds a second primer to target the site next to the padlock probe. \textbf{B) Sequencing-based methods.} This category provides an unbiased analysis of the complete transcriptome, capturing transcripts in situ and performing sequencing. First, the targeted tissue is placed on top of the microscopic slide, including a barcoded array that captures the spatial information related to each probe. Tiny needles inside each probe contain a spatial barcode and RT primers. After removing the tissue, cDNA-mRNA complexes are extracted for library preparation and next-generation sequencing (NGS) readout. Experimentally, the measured gene expressions are captured in spots or beads, complemented by a high-resolution histology image obtained by microscopy of stained tissue sections for the same tissue section. The dimensions of probes (spots/beads) vary, corresponding to the different technologies. They can be 100 µm (ST) or 55 µm (10X Visium) in diameter or use an ordered bead array onto which two µm-sized beads(HDST) or use ten µm-sized barcoded beads (Slide-seq).}
			\label{figure0}
		\end{figure*}

		In multicellular organisms, the tissues contain a group of diverse cells, with each cell conducting a specific function and constantly proliferating itself through the process of division \cite{larsson2021spatially}. A cell’s fate and behavior are associated with communicating to the surrounding environment. Awareness of the cell’s position and how it spatially organizes within a tissue is critical for understanding not only the function of the tissue, but also general concepts underlying diseases. For example, it has been shown that the cell-cell interactions between stromal and infiltrating immune cells play a major role in many physiological and pathological processes such as autoimmunity and cancer \cite{burkly2011tweak}. Single-cell RNA sequencing (scRNA-seq) has become a powerful approach
		in the genomics area, capturing the activity of thousands of genes in a biological sample at an unprecedented resolution. Currently, scRNA-seq has provided systematic benchmarks to dissect heterogeneous cell populations across various disciplines such as cancer, immunology, developmental biology, etc. \cite{suva2019single}. However, the scRNA-seq method requires tissue dissociation, leading to loss of cell position within the tissue, which is key to understanding the functionality of complex tissues. Spatially resolved transcriptomics (SRT), selected as method-of-the-year in 2020 \cite{xiaowei2021method}, has enabled researchers to capture the expression of genes along with corresponding spatial information across tissues \cite{crosetto2015spatially}.
		SRT data are generated based on the different experimental protocols. Generally, SRT technologies can be broadly  divided into two leading groups \cite{rao2021exploring} (I) image-based methods with high spatial resolution and overall low sensitivity in gene detection and (II) sequencing-based methods with limited spatial resolution but high-throughput mRNA-capturing. In  the first group, in-situ hybridization (ISH) methods enable the quantification of gene expression at a sub-cellular resolution and the visualization of RNA molecules directly in their original environment. These methods include spatially resolved transcript amplicon readout mapping (STARmap) \cite{starmap} and single-molecule fluorescent ISH (smFISH) \cite{smfish}. smFISH was further developed into the sequential hybridizations (seqFISH) \cite{seqfish}, multiplexed error-robust FISH (MERFISH) \cite{merfish}, and ouroboros smFISH (osmFISH) \cite{smfish} techniques, which each measure more mRNA species with higher resolution, respectively. Although current image-based methods can provide a higher gene detection sensitivity than sequencing-based methods, their resolution is inversely associated with the number of genes imaged, and they are limited to a specific number of preselected genes \cite{shengquan2021stplus}.
		Also, these methods are typically limited to hundreds of preselected genes. The sequencing-based methods depend on the prior spatial barcoding to perform an in situ capturing of transcripts \cite{asp2020spatially} followed by an in situ sequencing, such as spatial transcriptomics (ST)/10x-Visium, Slide-Seq, and high-definition spatial transcriptomics (HDST). This category empowers unbiased profiling of the complete transcriptome \cite{asp2020spatially}. 
		Therefore, it can capture thousands of genes at specific locations, denoted as spots at lower cellular resolution than image-based techniques. In the sequencing-based methods, hematoxylin and eosin (H\&E) stained histology images, provide necessary information about cellular morphology and heterogeneity of tissues in parallel \cite{zeng2022statistical}. Figure \ref{figure0} summarizes various SRT methods related to the image-based and sequencing-based approaches.\\ Existing histology images and spatial information generated using SRT methods and gene expression data have added a new dimension to omics research, generating massive and diverse datasets \cite{preibisch2021image}. Same as other biological data, SRT data are highly dimensional and implicitly noisy \cite{ij2018statistics}, which increases a demand for statistic and machine learning (ML) methods to deal with these challenges in SRT. Statistic approaches primarily focus on inference, meaning assuming a probability model on the input data to formalize understanding of a hypothesis about the system's behavior. In contrast, ML methods have a long-standing focus on prediction, in which the learning algorithms extract highly robust and rich features from data. Subsequently, some methods belong squarely to one domain or are common in both domains. However, the relationship between model complexity and the number of features (data-wide) and possible associations among them are linear, in which statistical inferences become less precise \cite{ij2018statistics}. Since the SRT data can be categorized as wide data in which the number of input variable are more than the number of observations, ML approaches can be more robust and efficient than statistical methods. As a robust ML approach, deep learning (DL) has proved its efficiency in many biological tasks (both supervised and unsupervised), particularly in the various steps of scRNA-seq data analysis such as normalization \cite{wang2019data}, dimensionality reduction \cite{lin2017using,eraslan2019single}, clustering \cite{tian2019clustering}, cell-type identification \cite{chung2017single}, data integration \cite{haghverdi2018batch,lopez2018deep,lotfollahi2022mapping} and perturbation modeling \cite{lotfollahi2020conditional,lotfollahi2019scgen,lotfollahi2021learning}.  However, SRT datasets are more challenging than scRNA-seq data due to their multi-modality and diversity. Consequently, conventional (ML) methods in the SRT data analysis are mainly similar to the statistical inference domain, in which there is a demand for pre-existing knowledge about the data to estimate unknown parameters in the model.
		Consequently, the DL method does not need to know the data-generation process to model data and is more potent in extracting complex and high-dimensional features. DL models are more versatile for integrating histology images, gene expression matrices, and spatial information. Indeed, DL paradigms have facilitated the handling of such complicated datasets and related downstream analyses. This paper undertakes a comprehensive review of recently developed ML and DL methods for analyzing SRT data (both imaging-based and sequencing-based techniques), focusing on DL models to investigate how DL approaches jointly use histology image, gene expression, and spatial coordinates, and how the spatial information can provide unprecedented insight into the molecular organization in heterogeneous cellular contexts. Some efforts have been made to review the computational challenges in the SRT domain. Hu et al. \cite{hu2021statistical}, focused on the statistical and ML methods to analyze SRT data. This work focused on the capacity of histology images which can be applied to both imaging-based and sequencing-based techniques. Zeng et al. \cite{zeng2022statistical}, provided a summary of the statistical and ML methods in the SRT domain with more focus on the sequencing-based methods. Despite the valuable information in these review papers, they did not provide detailed information and discussion of the application of DL models in SRT analysis. Although the review paper by Heydari and Sindi \cite{heydari2022deep} reviewed the application of DL in SRT analysis, this work mainly focused on sequencing-based approaches. In this technical review, we identified papers published on applying ML methods focusing on Dl models for analyzing both imaging- and sequencing-based SRT data up to June 2022. We summarize the concepts and common tasks in SRT data analysis into six categories and discuss the DL models and their associated findings in detail.
		Our study also offers complete information on current SRT datasets and evaluation metrics. Also, we provide the technical detail of current DL models and the associated results in the supporting material. We envisage that this paper will serve as a comprehensive reference for further applications of DL in SRT
		data analysis and can facilitate the development of innovative methods in the future. 
		
		\section{Overview of common deep learning models for SRT data analysis}
\begin{figure*}
			\centering
			\begin{adjustbox}{max width=\textwidth}
				\begin{tikzpicture}
					\draw [gray,thick,fill=blue!20](4,28) rectangle (10,30);
					\draw [gray,thick,fill=blue!20](11,28) rectangle (17,30);
					\draw [gray,thick,fill=blue!20](18,28) rectangle (24,30);
					\node at (7,29) [font=\sffamily] {\large Gene Expression Matrix};
					\node at (14,29) [font=\sffamily] {\large Spatial Coordinates};
					\node at (21,29) [font=\sffamily] {\large Histology Image};
					\draw [gray,thick,rounded corners,fill=green!10](0,21) rectangle 				(2,27);
					\node at (1,24)[rotate=90] [font=\sffamily] {\huge 							Pre-processing};
					\draw [thick,red!80,dashed](4,21) rectangle 									(10,27);
					\draw [red!80,thick,->] (7,28)--(7,27);
					\draw [thick,green!80,dashed](11,21) rectangle 									(17,27);
					\draw [green!80,thick,->] (14,28)--(14,27);
					\draw [thick,blue!80,dashed](18,21) rectangle 									(24,27);
					\draw [blue!80,thick,->] (21,28)--(21,27);
					\draw [fill] (4.2,26.7) circle (.1cm);
					\node at (4.4,26.7) [font=\sffamily,right] {\large Eliminating 					genes that are };
					\node at (4.4,26) [font=\sffamily,right] {\large expressed less 				than $x$ spots};
					\draw [fill] (4.2,25) circle (.1cm);
					\node at (4.4,25) [font=\sffamily,right] {\large Dimension 					Reduction};
					\draw [fill] (4.2,24) circle (.1cm);
					\node at (4.4,24) [font=\sffamily,right] {\large Normalization};
					\draw [fill] (5,23) circle (.1cm);
					\node at (5.2,23) [font=\sffamily,right] {\large log-							transformation};
					\draw [fill] (5,22) circle (.1cm);
					\node at (5.2,22) [font=\sffamily,right] {\large Z-								transformation};
					\draw [fill] (11.2,26.7) circle (.1cm);
					\node at (11.4,26.7) [font=\sffamily,right] {\large Constructing 				Adjacent matrix};
					\node[scale=.3,red,draw,circle,label=above:$g_{4}$] (g4) at 					(12,25){} ;
					\node[scale=.3,red,draw,circle,label=above:$g_{2}$] (g2) at 					(13,25){} ;
					\node[scale=.3,red,draw,circle,label=above:$g_{3}$] (g3) at 					(12.5,25.5){} ;
					\node[scale=.3,red,draw,circle,label=above:$g_{1}$] (g1) at 					(13.5,25){} ;
					\node[scale=.3,red,draw,circle,label=above:$g_{5}$] (g5) at 					(15,25){} ;
					\node[scale=.3,red,draw,circle,label=above:$g_{6}$] (g6) at 					(12.5,24.5){} ;
					\node[scale=.3,red,draw,circle,label=above:$g_{7}$] (g7) at 					(14,24){} ;
					\node[scale=.3,red,draw,circle,label=above:$g_{8}$] (g8) at 					(14,26){} ;
					\draw [dotted] (13.5,25) circle (1.6cm);
					\draw [red,dashed] (13.5,25) circle (1cm);
					\draw [violet,thick,-] (13.5,25)--(14.5,25);
					\node at (14,25.2) [font=\sffamily,violet] {\large $R$};
					\node at (11.5,25) [rotate=90,font=\sffamily] {\large Spot};
					\draw [thick,-] (13,23)--(13,21);
					\draw [thick,-] (16,23)--(16,21);
					\node at (12.1,22) [right,font=\sffamily] {\large A=};
					\node at (12.5,22.8) [right,font=\sffamily] {\large $g_1$};
					\node at (13,23.2) [right,font=\sffamily] {\large $g_1$};
					\node at (13.5,23.2) [right,font=\sffamily] {\large $g_2$};
					\node at (15.4,23.2) [right,font=\sffamily] {\large $g_8$};
					\node at (14,23.2) [right,font=\sffamily] {\large ....};
					\node at (13,22.8) [right,font=\sffamily] {\large 1};
					\node at (13.5,22.8) [right,font=\sffamily] {\large 1};
					\node at (15.4,22.8) [right,font=\sffamily] {\large 0};
					\node at (12.8,22.4) [rotate=90,font=\sffamily] {\large ....};
					\node at (21,23.5) {\includegraphics[scale=0.2]{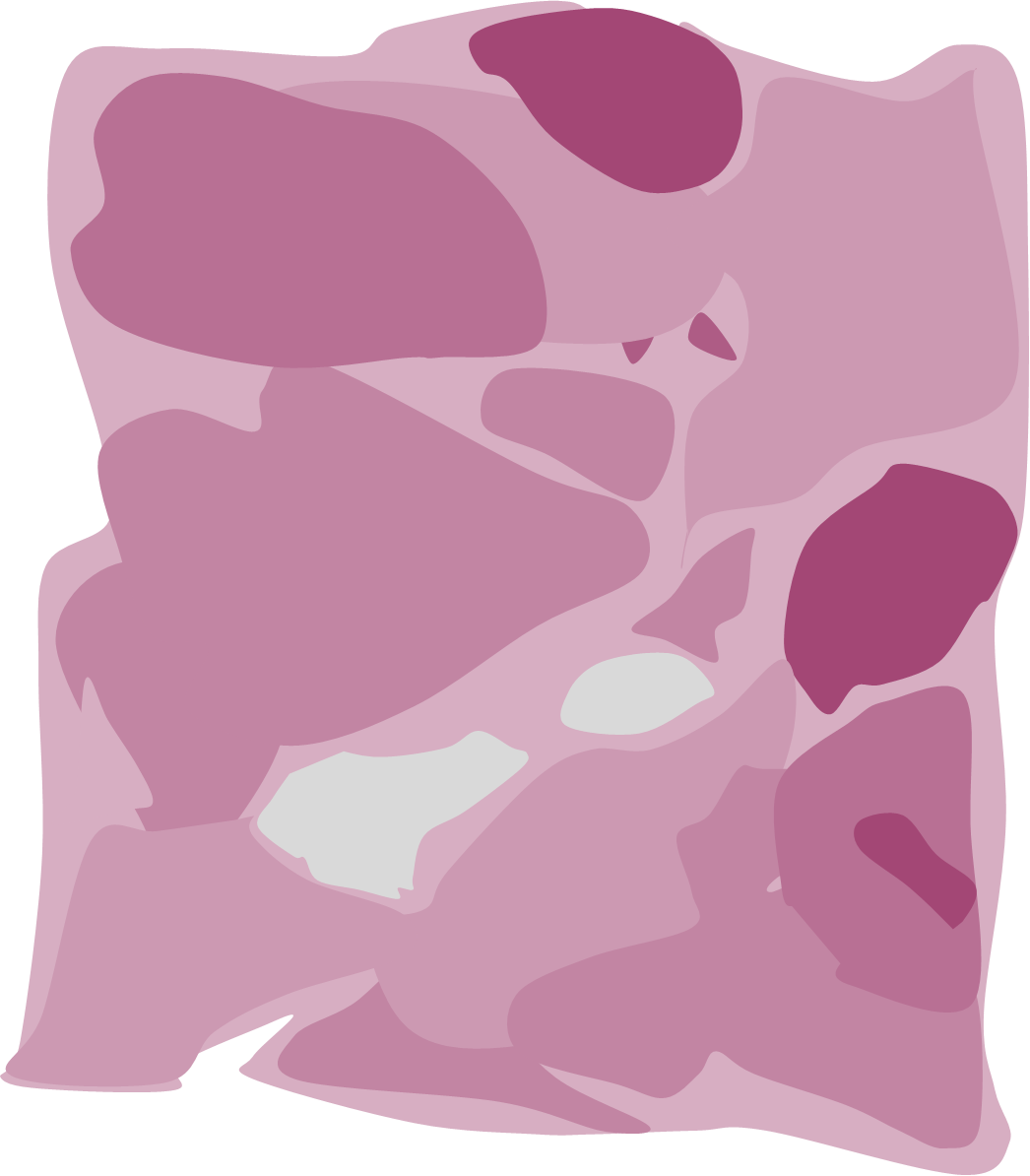}};
					\draw [thin,step=0.4] (19.2,21.5) grid (22.8,25.7);
					\draw [fill] (18.2,26.7) circle (.1cm);
					\node at (18.4,26.7) [right,font=\sffamily] {\large Tiling the 					input image into the};
					\node at (18.4,26.2) [right,font=\sffamily] {\large patches};
					\draw [thick](3,11) rectangle (7,20);
					\draw [thick](7.5,11) rectangle (11.5,20);
					\draw [thick](12,11) rectangle (16,20);
					\draw [thick](16.5,11) rectangle (20.5,20);
					\draw [thick](21,11) rectangle (25.2,20);
					\draw [gray,thick,rounded corners,fill=green!10](0,11) rectangle 				(2,20);
					\node at (1,15.5) [rotate=90,font=\sffamily] {\huge Deep 						Learningg Models};
					\node at (5,19.5) [font=\sffamily] {\huge DNN };
					\node[circle,draw, minimum size = 6mm]  at (6.5,19.5) {4};
					\node at (9.5,19.5) [font=\sffamily] {\huge AE };
					\node[circle,draw, minimum size = 6mm]  at (11,19.5) {6};
					\node at (14,19.5) [font=\sffamily] {\huge VAE };
					\node[circle,draw, minimum size = 6mm]  at (15.5,19.5) {4};
					\node at (18.5,19.5) [font=\sffamily] {\huge CNN };
					\node[circle,draw, minimum size = 6mm]  at (20,19.5) {11};
					\node at (23,19.5) [font=\sffamily] {\huge GCN };
					\node[circle,draw, minimum size = 6mm]  at (24.5,19.5) {7};
					\node at (5,18) [font=\sffamily] {\large ....};
					\newcommand{\inputnum}{2} 
					\newcommand{\hiddennum}{4}  
					\newcommand{\picnum}{4}
					\newcommand{\outputnum}{1} 
					\foreach \i in {2,4}
					{
						\node[circle, 
						minimum size = 6mm,
						fill=red!50] (Input+\i) at (2+\i,18) {};
					}
					\foreach \i in {1,2,3,4}
					{
						\node[circle, 
						minimum size = 6mm,
						fill=blue!50] (Hidden1+\i) at (2.5+\i,16) {};
					}
					\foreach \i in {1,2,3,4}
					{
						\node[circle, 
						minimum size = 6mm,
						fill=orange!50] (Hidden2+\i) at (2.5+\i,14) {};
					}
					\foreach \i in {2,4}
					{
						\foreach \j in {1,...,\hiddennum}
						{
							\draw[->, shorten >=1pt] (Input+\i) -- (Hidden1+\j);   
						}
					}
					\foreach \i in {1,...,\hiddennum}
					{
						\foreach \j in {1,...,\hiddennum}
						{
							\draw[->, shorten >=1pt] (Hidden1+\i) -- (Hidden2+\j);   
						}
					}
					\node[circle, 
					minimum size = 6mm,
					fill=gray] (out) at (5,12.5) {$\sum$};
					\foreach \i in {1,...,\hiddennum}
					{
						\draw[->, shorten >=1pt] (Hidden2+\i) -- (out);    
					}
					\node[trapezium,draw= blue!80,font=\sffamily,fill=orange!30,rotate=-180,scale=1.5] (t) at (9.5,16.5) {\rotatebox{180}{Encoder}};
					\node[rectangle,draw=blue!80,fill=orange!10,minimum width = 3cm, 
					minimum height = 0.5cm,scale=1](t0) at (9.5,18) {$X$};
					\draw[->, shorten >=1pt] (t0) -- (t);
					\node[rectangle,draw=blue!80,fill=orange!10,minimum width = 3cm, 
					minimum height = 0.75cm,scale=1,text = red,font=\sffamily](ls) at (9.5,15) {Latent Space $Z$};
					\draw[->, shorten >=1pt] (t) -- (ls);
					\node[trapezium,draw= blue!80,font=\sffamily,fill=orange!30,scale=1.5] (d) at (9.5,13.5) {Decoder};
					\node[rectangle,draw=blue!80,fill=orange!10,minimum width = 3cm, 
					minimum height = 0.5cm,scale=1](t1) at (9.5,12) {$\hat X$};
					\draw[->, shorten >=1pt] (ls) -- (d);
					\draw[->, shorten >=1pt] (d) -- (t1);
					\node[trapezium,draw= blue!80,font=\sffamily,fill=orange!30,rotate=-180,scale=1.5] (f) at (14,16.5) {\rotatebox{180}{$Q(Z \mid X)$}};
					\node[rectangle,draw=blue!80,fill=orange!10,minimum width = 3cm, 
					minimum height = 0.5cm,scale=1](f0) at (14,18) {$X$};
					\draw[->, shorten >=1pt] (f0) -- (f);
					\node[rectangle,draw=blue!80,fill=orange!10,minimum width = 3cm, 
					minimum height = 0.75cm,scale=1,text = red,font=\sffamily](ls2) at (14,15) {Latent Space $Z$};
					\draw[->, shorten >=1pt] (f) -- (ls2);
					\node[trapezium,draw= blue!80,font=\sffamily,fill=orange!30,scale=1.5] (d2) at (14,13.5) {$P(Z \mid X)$};
					\node[rectangle,draw=blue!80,fill=orange!10,minimum width = 3cm, 
					minimum height = 0.5cm,scale=1](f1) at (14,12) {$\hat X$};
					\draw[->, shorten >=1pt] (ls2) -- (d2);
					\draw[->, shorten >=1pt] (d2) -- (f1);
					\foreach \x in {1,...,4} {
						\begin{scope}[cm={1,0.3,0,1,(17+\x-1,18)}]
							\node[transform shape, draw, red, ultra thick, inner sep=0.2mm] {
								\includegraphics[width=1cm]{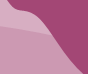}
							};
					\end{scope}}
					\tikzset{my_rectangle/.style={draw,rectangle,fill=violet!50, minimum width=2cm, minimum height=1cm}}
					\tikzset{my_rectangle2/.style={draw,rectangle,fill=yellow!50, minimum width=1cm, minimum height=.5cm}}
					\foreach \i in {0,.5,1,1.5,2,2.5,3} {
						\node[my_rectangle,canvas is zy plane at x=1 ]at (16+\i,16) {};}
					\draw[->, shorten >=1pt] (20,17.5) -- (20,17);
					\node at (16.5,19) [right,font=\sffamily] {Input patches };
					\node at (16.5,17) [right,font=\sffamily] {Convolution layer};
					\foreach \i in {0,.5,1,1.5,2,2.5,3} {
						\node[my_rectangle2,canvas is zy plane at x=1 ]at (16+\i,14) {};}
					\draw[->, shorten >=1pt] (20,15) -- (20,14.5);
					\node at (16.5,14.7) [right,font=\sffamily] {Pooling layer};
					\draw[->, shorten >=1pt] (20,13.5) -- (20,13);
					\foreach \i in {1,2,3}
					{
						\node[circle, 
						minimum size = 5mm,
						fill=orange!50] (fc+\i) at (16.5+\i,13) {};
					}
					\node at (16.5,13) [right,font=\sffamily] {FC};
					\node[circle, 
					minimum size = 6mm,fill=gray] (o) at (18.5,12) {$\sum$};
					\foreach \i in {1,2,3}
					{
						\draw[->, shorten >=1pt] (fc+\i) -- (o);    
					}  
					\begin{scope}[every node/.style={circle,thick,draw,scale=.7,fill=blue!40}]
						\node (g1) at (21.5,17) {$g_1$};
						\node (g2) at (22,18) {$g_2$};
						\node (g3) at (23,18.5) {$g_3$};
						\node (g4) at (24,18) {$g_4$};
						\node (g5) at (23,17) {$g_5$};
						\node (g6) at (24,17) {$g_6$} ;
					\end{scope}
					\draw (g1) -- (g2);\draw (g2) -- (g3);\draw (g3) -- (g4);\draw (g4) -- (g6);
					\draw (g6) -- (g5);\draw (g5) -- (g2);
					\begin{scope}[every node/.style={circle,thick,draw,scale=.4,fill=blue!40}]
						\node (g11) at (23.5,15) {$g_1$};
						\node (g22) at (24,15.5) {$g_2$};
						\node (g33) at (24.5,16) {$g_3$};
						\node [red,text=red](g44) at (25,15.5) {$g_4$};
						\node (g55) at (24.5,15) {$g_5$};
						\node (g66) at (25,15) {$g_6$} ;
					\end{scope}
					\draw (g11) -- (g22);\draw (g22) -- (g33);\draw[red] (g33) -- (g44);\draw[red] (g44) -- (g66);
					\draw (g66) -- (g55);\draw (g55) -- (g22);
					\begin{scope}[every node/.style={circle,thick,draw,scale=.4,fill=blue!40}]
						\node (g111) at (21.5,15) {$g_1$};
						\node (g222) at (22,15.5) {$g_2$};
						\node[red,text=red] (g333) at (22.5,16) {$g_3$};
						\node (g444) at (23,15.5) {$g_4$};
						\node (g555) at (22.5,15) {$g_5$};
						\node (g666) at (23,15) {$g_6$} ;
					\end{scope}
					\draw (g111) -- (g222);\draw[red]  (g222) -- (g333);\draw[red] (g333) -- (g444);\draw(g444) -- (g666);
					\draw (g666) -- (g555);\draw (g555) -- (g222);
					\node at (21.5,15.5) {\huge ...} ;
					\foreach \i in {1,2,3}
					{
						\node[circle, 
						minimum size = 5mm,
						fill=orange!50] (fc+\i) at (21+\i,14) {};
					}
					\begin{scope}[every node/.style={circle,thick,draw,scale=.7,fill=blue!40}]
						\node[olive] (gg1) at (21.5,11.25) {$g_1$};
						\node [olive] (gg2) at (22,12.25) {$g_2$};
						\node[olive]  (gg3) at (23,12.75) {$g_3$};
						\node[yellow]  (gg4) at (24,12.25) {$g_4$};
						\node [yellow] (gg5) at (23,11.25) {$g_5$};
						\node [yellow] (gg6) at (24,11.25) {$g_6$} ;
					\end{scope}
					\draw (gg1) -- (gg2);\draw (gg2) -- (gg3);\draw (gg3) -- (gg4);\draw (gg4) -- (gg6);
					\draw (gg6) -- (gg5);\draw (gg5) -- (gg2);
					\draw[->, shorten >=1pt] (24.5,17) -- (24.5,16.75);
					\node at (21,16.5) [right,font=\sffamily] {Convolution layer};
					\draw[->, shorten >=1pt] (24,14.75) -- (24,14.25);
					\draw[->, shorten >=1pt] (23,14.75) -- (23,14.25);
					\draw[->, shorten >=1pt] (22,14.75) -- (22,14.25);
					\node at (21,14) [right,font=\sffamily] {FC};
					\node at (20.9,13.2) [right,font=\sffamily] {Classification};
					\draw[->, shorten >=1pt] (23,13.7) -- (23,13.2);
					
				\end{tikzpicture}
			\end{adjustbox}
			\caption{\textbf{Graphical models of the surveyed deep learning models in spatially resolved transcriptomics.} These methods include deep neural network (DNN), deep autoencoder (AE), variational autoencoder (VAE), convolutional neural network (CNN); and graph convolution network (GCN). The number beside each model's name refers to the number of reviewed papers that employed this specific deep model till June 2022.
			}
			\label{figure1}
		\end{figure*}

		According to the existing tasks in SRT exploration, the reviewed papers used different supervised and unsupervised learning methods in their works. For example, gene prediction, and cell segmentation are supervised, whereas clustering, imputation, and dimension reduction are unsupervised learning tasks. For a better understanding of the reviewed methods in this study, we first describe the DL models that have been used to analyze SRT data as well as their general mathematical formula together with their training strategies. These models include deep neural networks (DNNs), autoencoders (AEs), variational autoencoders (VAEs), convolutional neural networks (CNNs), and graph convolution networks (GCNs) \cite{goodfellow2016deep}. We focus on different attributes of each method along with their uniqueness and novelty. Figure \ref{figure1} illustrates a brief visualization of all surveyed DL models in this paper along with available pre-processing approaches regarding gene expression matrices, spatial information, and histology images. The DL model and the pre-processing steps vary depending on the input data and the research objectives.
		\subsection{Deep Neural Networks}
		\label{dnn} 
		The DNNs, called feed-forward neural networks, are the quintessential DL model. These networks consist of multiple hidden layers (fully connected layers), each of which including several neurons. The number of layers determines the depth of the model. Given the input $X$, the DNN model approximates the nonlinear transformation $\phi$ for a specific goal (e.g., classification). The neurons of each layer ($l$), take the output of neurons in the previous layer ($l-1$) and feed them to an activation function $y_n=f(x_{n-1}.w_{in}+b)$, where $w_n$ is the weight of neuron $i$, and $b$ is the bias. The weights and biases in each layer are the parameters that are updated using a predefined optimization algorithm that minimizes a loss function.
		\subsection{Autoencoders}
		\label{AE}
		The AE networks are deep generative models, mostly considered as a dimensionality reduction method. The AEs have an encoder part $E$ and a decoder part $D$, with specific neural network architectures in each part, respectively. The encoder part takes the input data $X$ and generates the latent variable $Z$ by learning the network weights $\theta$ as $Z=E_{\theta}(X)$.
		The latent representation $Z$ carries out the most important information of the input, which can be leveraged in clustering and other downstream analyses. The decoder part receives $Z$ as the input and reconstructs the input data $X$ by updating the decoder network weights $\phi$  during the training as $\hat{X}=D_{\phi}(Z)$.
		All network parameters ($\theta$ and $\phi$) are updated by minimizing the mean square error (MSE) loss function between the input data $X$ and reconstructed input $\hat{X}$.
		The goal of AE models is to introduce a paired encoder-decoder that keep the maximum information when encoding and then achieve the minimum of reconstruction error in the decoding.  
		\subsection{Variational Autoencoders}
		\label{VAE}
		The AEs encode an input as a single point without any regularity in the latent space, leading to a lack of generative properties in the decoder part. Variational autoencoders (VAEs) are deep generative models which encode the input as a distribution over the latent space.
		They consist of two parts: inference and generative model. The inference model, also called the recognition model or encoder takes the input data $X$ and learns the latent features $Z$ through the multi-layer deep neural networks. In the inference process, there are two assumptions: (1) the latent variable $Z$ is sampled from a prior distribution $p(Z)=N(Z;0, I)$, and (2) the input data $X$ follows the assumed conditional distribution (e.g., Gaussian distribution with mean $\mu$ and variance $\sigma$) $q_\theta (Z\mid X)-N(Z;\mu,\sigma)$, where $\theta$ denoted the encoder network's weight. The decoder part is a Bayesian network, which accepts the variable $Z$ and calculates the posterior probability $P_\phi (X, Z)$, where $\phi$ is the decoder network's weights. Therefore, the main aim of the VAE is to find the optimal value for the distribution parameters by updating the $\theta$ and $\phi$, to maximize the evidence lower bound (ELBO) as follows:
		\begin{equation}
			\label{elbo}
			L_{\theta,\phi}(x)=\log(p_\phi (x))-D_{KL}(q_\theta (Z \mid X) \parallel P_\phi (Z \mid X))
		\end{equation} 
		where, $\log(p_\phi (x))$ is the marginal likelihood and $D_{KL}$ is the Kullback-Leibler (KL) divergence between the approximates and true (posterior) distribution. Eq.\ref{elbo} is optimized by the stochastic gradient descent (SGD) approach. The VAE can be applied to graphs, giving rise to the variational graph autoencoder (VGAE). Besides this general formulation (Eq.\ref{elbo}), each paper proposed a specific data distribution and learning strategy.
		\subsection{Convolutional Neural Networks}
		\label{CNN}
		CNNs are well-known supervised methods that have proved their efficiency in many areas, especially in image processing. Briefly, a CNN has multiple layers, such as convolutional, pooling, normalization, dropout, and fully connected. A multidimensional array of data is usually fed as the input to perform feature extraction. In the convolution layer, the convolution operation is applied to the input matrix $X$ using the kernel $W$, i.e. $S=X \ast W$, where $S$ is often referred to as the feature map. An essential attribute of the convolutional layer is the commutative property, which arises because the kernel is flipped relative to the input to learn the most important features (e.g., edges in the image) regarding their locations. Then, the output is passed to the non-linear activation function. Typically, the output of the activation function is run through the pooling layer to reduce the dimensions of the input layer. The pooling operation such as max-pooling or average-pooling replaces the previous layers' output at a particular location with a maximum or mean of a rectangular
		neighborhood. Mainly, the pooling layer helps to make the representation approximately
		invariant to small input translations meaning small changes in the input do not result in large changes in the output. CNNs are used more than other DL models on SRT data for gene classification, gene prediction, and learning embedded features because of existing histology images. The details of the CNN models used in the reviewed papers are discussed later. 
		\subsection{Graph Convolutional Networks}
		\label{GCN}
		The idea of GCNs is the same as the CNNs, but the GCNs take the graphs (irregular data) as the input, and the kernels fit on each node and their adjacent neighbors. Given the input graph $G(E,V)$ with $N$ vertices $v_i\in V$, edges $(v_i,v_j\in E)$, an adjacent matrix $A\in R^{N\times N}$, degree matrix $D_{ii}=\sum_{j}A_{ij}$, and input matrix $X\in R^{N\times C}$, where $C$ is the dimension of the feature maps. The network gathers the information from a given vertex’s neighbors and transfers them to the next layer according to the value of each vertex’s features. Thus, selecting the appropriate layer number allows the network to learn the graph structure to perform various tasks, such as graph and vertex classification, vertices and link prediction, and clustering. Since the SRT data comprises spatial information, the graph with genes or cells (in imaging-based technique) and spots (sequencing-based approaches) as a node can be structured for clustering and predicting unmeasured genes. Moreover, these models are helpful to find a reciprocal link between SRT and scRNA-seq data. If the GCN leverages a masked self-attention mechanism to learn weights between each pair of connected vertices, the network is named graph attention network (GAT). There are five GCN models (including one GAT model) in the reviewed papers, which are elaborated on later in this paper. \\
		Above mentioned models can be categorized as sequential models, generating a sequence of hidden states as a function of the previous hidden state. The problem of sequential mechanism is hindering parallelization within training examples, which becomes critical at longer
		sequence lengths in larger data sizes, leading to memory constraints \cite{vaswani2017attention}. Recently, attention mechanisms (HA) have been developed to reduce the restriction of sequential computation by designing dependencies without considering their distance in the input or output sequences. 
		Additionally, the multi-head attention mechanism (MHA) or transformer \cite{vaswani2017attention} is a robust model architecture, allowing more parallelization in deep neural network-based methods. Since the reviewed papers propose different architectures of DL models and various loss functions, we will explain their topologies in the next section \ref{survay}.

		\section{Survey of deep learning models for SRT analysis}
		\label{survay} 
\begin{figure*}
			\centering
			\begin{adjustbox}{max width=\textwidth}
				\begin{tikzpicture}
					\draw [rounded corners,dashed,thick](3,19.5) rectangle (22.5,27.5);
					\node at (6,24) {\includegraphics[scale=0.3]{org.png}};
					\node at(5.5,20.5){{\large \textbf{Histology Image}\par}};
					\draw [fill,color=myblue] (7.5,20.5) circle (.25cm);
					\node at (14.1,23.8) {\includegraphics[scale=0.7]{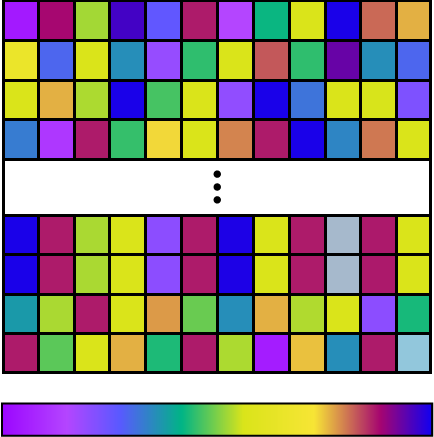}};
					\node at(10.5,26.2){{\small Spot/Cell 1\par}};
					\node at(10.5,25.6){{\small Spot/cell 2\par}};
					\node at(11,25.3){{\large .\par}};
					\node at(11,24.8){{\large .\par}};
					\node at(12,27)[rotate=45]{\small{Gene 1}};
					\node at(12.4,27)[rotate=45]{\small{Gene 2}};
					\node at(12.5,26.7)[rotate=45]{\small{.}};
					\node at(13,26.7)[rotate=45]{\small{.}};
					\draw (19.5,22) rectangle (21.5,26.4);
					\node at(20,26.6){$X$};\node at(21,26.6){$Y$};
					\node at(20,26.1){$x_1$};\node at(21,26.1){$y_1$};
					\node at(20,25.6){$x_2$};\node at(21,25.6){$y_2$};
					\node at(20,25.2){.};\node at(21,25.2){.};
					\node at(20,24.8){.};\node at(21,24.8){.};
					\node at(14,20.5){{\large \textbf{Gene expression}\par}};
					\node at(19.8,20.5){{\large\textbf{Spatial coordinates}\par}}; 
					\draw [fill,color=myred] (16,20.5) circle (.25cm);
					\draw [fill,color=mynaiveblue] (22,20.5) circle (.25cm);
					\draw [dashed] (8.2,26.2) circle (.2cm);
					\draw [thick,->](8.5,26.2)--(9.5,26.2);
					\draw [thick](20.5,26.4)--(20.5,22);
					\draw [gray,rounded corners,dashed,thick,fill=yellow!20](2.5,12.5) rectangle (11.5,18);
					\node at(7,17.5){{\large Identifying Spatial Domain\par}};
					\node at (4.25,15.1) {\includegraphics[scale=0.2]{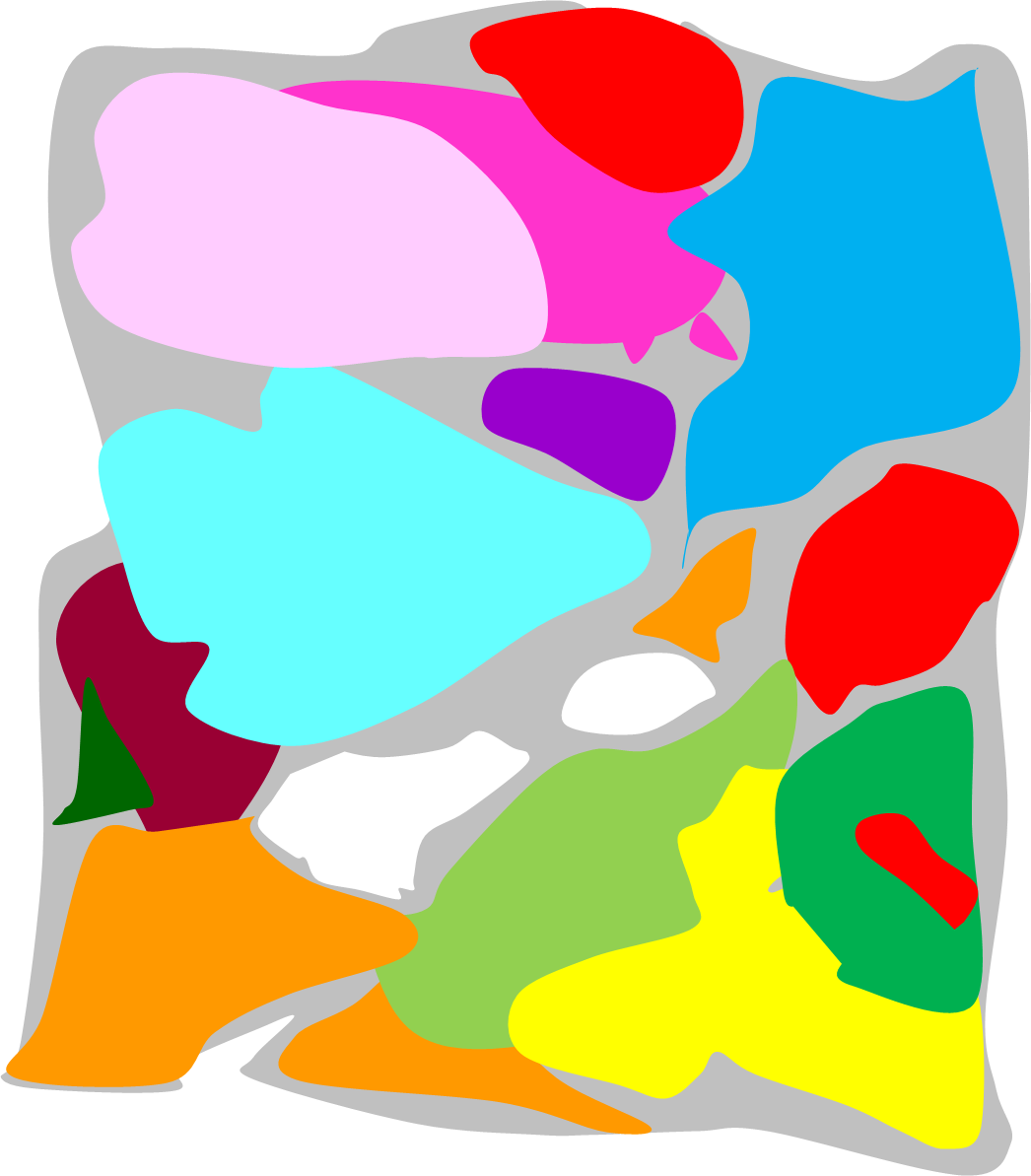}};
					\node[right] at(6,17){{\large SpaCell\par}};\node[right] at(8.5,17){{\large STAGATE\par}};
					\node[right] at(6,16){{\large stLearn\par}};\node[right] at(8.5,16){{\large RESEPT\par}};
					\node [right]at(6,15){{\large SpaGCN\par}};\node[right] at(8.5,15){{\large ECNN\par}};
					\node[right] at(6,14){{\large SEDR\par}};\node[right] at(8.5,14){{\large JSTA\par}};
					\node[right] at(6,13){{\large conST\par}};
					\coordinate (center) at (8,17);
					\fill[myblue] (center) + (0, 0.25) arc (90:270:.25);
					\fill[myred] (center) + (0, -0.25) arc (270:450:0.25);
					\tkzDefPoint(8,16){O}
					\foreach \mycolor/\mygrad in {myred/0,myblue/120,mynaiveblue/240}
					\tkzDrawSector[R,draw=white,fill=\mycolor](O,.25)(\mygrad,\mygrad+120);
					\tkzDefPoint(8,15){O}
					\foreach \mycolor/\mygrad in {myred/0,myblue/120,mynaiveblue/240}
					\tkzDrawSector[R,draw=white,fill=\mycolor](O,.25)(\mygrad,\mygrad+120);
					\coordinate (center) at (8,14);
					\fill[mynaiveblue] (center) + (0, 0.25) arc (90:270:.25);
					\fill[myred] (center) + (0, -0.25) arc (270:450:0.25);
					\coordinate (center) at (10.75,17);
					\fill[mynaiveblue] (center) + (0, 0.25) arc (90:270:.25);
					\fill[myred] (center) + (0, -0.25) arc (270:450:0.25);
					\coordinate (center) at (10.75,16);
					\fill[mynaiveblue] (center) + (0, 0.25) arc (90:270:.25);
					\fill[myred] (center) + (0, -0.25) arc (270:450:0.25);
					\draw [fill,color=myblue] (10.75,15) circle (.25cm);
					\coordinate (center) at (10.75,14);
					\fill[myblue] (center) + (0, 0.25) arc (90:270:.25);
					\fill[myred] (center) + (0, -0.25) arc (270:450:0.25);
					\tkzDefPoint(8,13){O}
					\foreach \mycolor/\mygrad in {myred/0,myblue/120,mynaiveblue/240}
					\tkzDrawSector[R,draw=white,fill=\mycolor](O,.25)(\mygrad,\mygrad+120);
					\draw [gray,rounded corners,dashed,thick,fill=yellow!20](12,12.5) rectangle (22,18);
					\node at(17,17.5){{\large Identifying SVG\par}};
					\node at (14.3,15) {\includegraphics[scale=0.12]{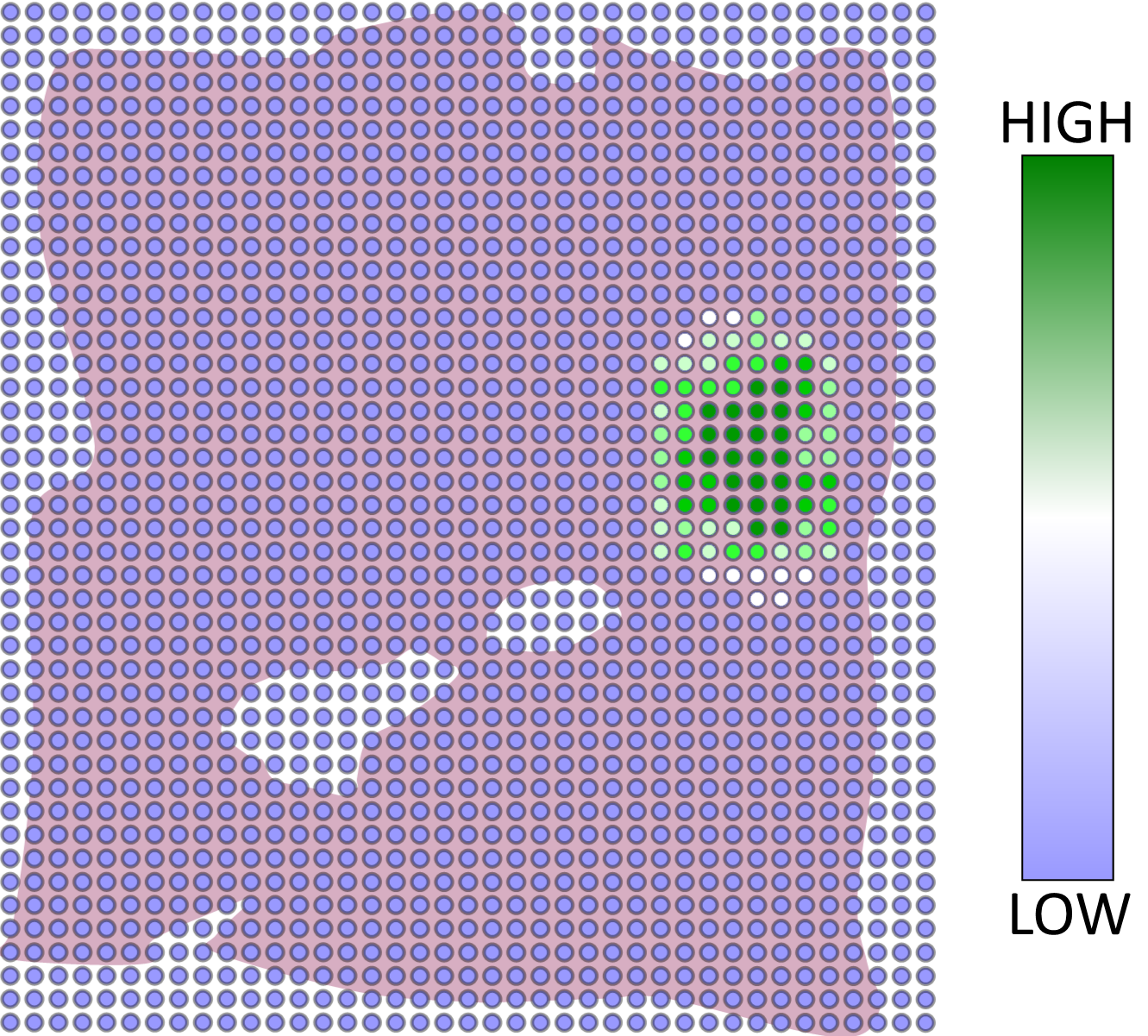}};
					\node[right] at(16.3,17){{\large SpaGCN\par}};\node[right] at(16.3,16){{\large STAGATE\par}};
					\node [right]at(16.3,15){{\large CoSTA\par}};\node[right] at(16.3,14){{\large SPADE\par}};
					\node[right] at(16.3,13){{\large conST\par}};
					\node[right] at(19,17){{\large ST-Net\par}};\node[right] at(19,16){{\large HisToGene\par}};
					\node[right] at(19,15){{\large CNNTL\par}};\node [right]at(19,14){{\large DeepSpaCE\par}};
					\tkzDefPoint(18.7,17){O}
					\foreach \mycolor/\mygrad in {myred/0,myblue/120,mynaiveblue/240}
					\tkzDrawSector[R,draw=white,fill=\mycolor](O,.25)(\mygrad,\mygrad+120);
					\coordinate (center) at (18.7,16);
					\fill[mynaiveblue] (center) + (0, 0.25) arc (90:270:.25);
					\fill[myred] (center) + (0, -0.25) arc (270:450:0.25);
					\tkzDefPoint(18.7,15){O}
					\foreach \mycolor/\mygrad in {myred/0,myblue/120,mynaiveblue/240}
					\tkzDrawSector[R,draw=white,fill=\mycolor](O,.25)(\mygrad,\mygrad+120);
					\coordinate (center) at (18.7,14);
					\fill[mynaiveblue] (center) + (0, 0.25) arc (90:270:.25);
					\fill[myred] (center) + (0, -0.25) arc (270:450:0.25);
					
					\coordinate (center) at (21.5,17);
					\fill[myblue] (center) + (0, 0.25) arc (90:270:.25);
					\fill[myred] (center) + (0, -0.25) arc (270:450:0.25);
					\draw [fill,color=myblue] (21.5,15) circle (.25cm);
					\tkzDefPoint(21.5,16){O}
					\foreach \mycolor/\mygrad in {myred/0,myblue/120,mynaiveblue/240}
					\tkzDrawSector[R,draw=white,fill=\mycolor](O,.25)(\mygrad,\mygrad+120);
					\coordinate (center) at (21.5,14);
					\fill[myblue] (center) + (0, 0.25) arc (90:270:.25);
					\fill[myred] (center) + (0, -0.25) arc (270:450:0.25);
					\tkzDefPoint(18.7,13){O}
					\foreach \mycolor/\mygrad in {myred/0,myblue/120,mynaiveblue/240}
					\tkzDrawSector[R,draw=white,fill=\mycolor](O,.25)(\mygrad,\mygrad+120);
					\draw [gray,rounded corners,dashed,thick,fill=yellow!20](0,6) rectangle (7,11.5);
					\node at(3.5,11){{\large Enhancing GER \par}};
					\node at (2,8.5) {\includegraphics[scale=.25]{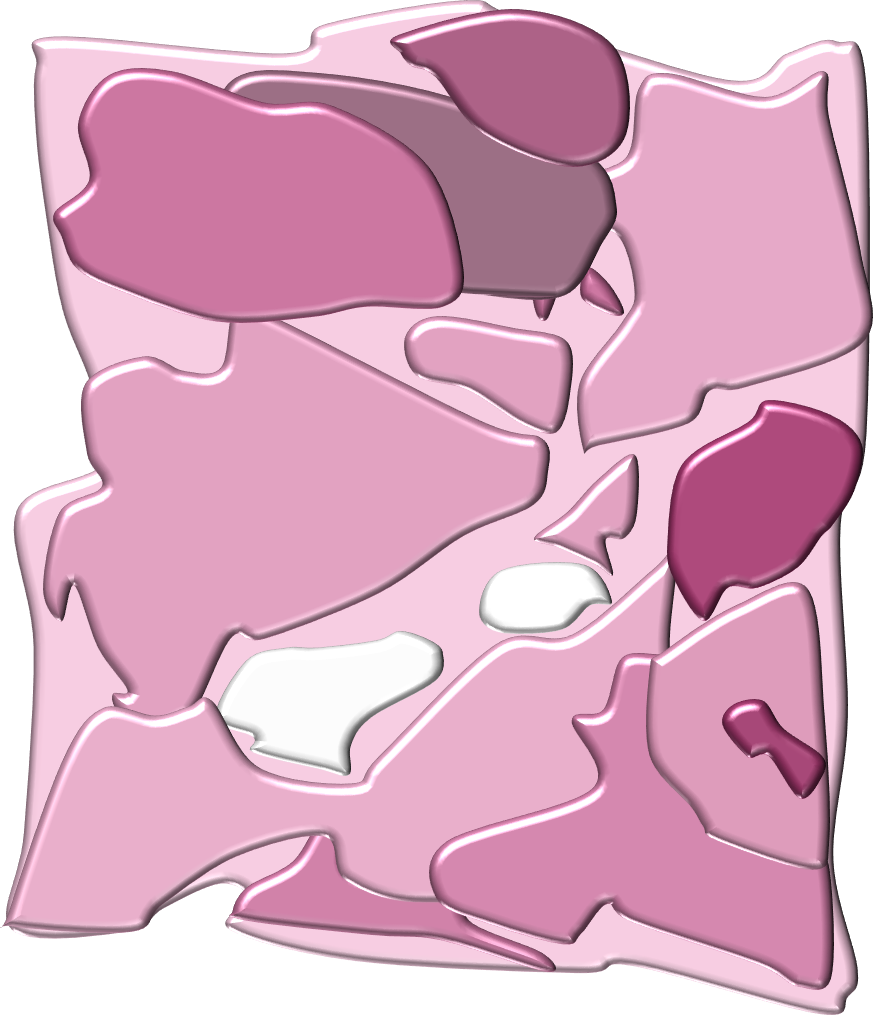}};
					\node[right] at(3.9,9.5){{\large XFuse\par}};
					\node [right]at(3.9,8.5){{\large DeepSpaCE \par}};
					\node[right] at(3.9,7.5){{\large HisToGene* \par}};
					\coordinate (center) at (6.5,9.5);
					\fill[myblue] (center) + (0, 0.25) arc (90:270:.25);
					\fill[myred] (center) + (0, -0.25) arc (270:450:0.25);
					\coordinate (center) at (6.5,8.5);
					\fill[myblue] (center) + (0, 0.25) arc (90:270:.25);
					\fill[myred] (center) + (0, -0.25) arc (270:450:0.25);
					\tkzDefPoint(6.5,7.5){O}
					\foreach \mycolor/\mygrad in {myred/0,myblue/120,mynaiveblue/240}
					\tkzDrawSector[R,draw=white,fill=\mycolor](O,.25)(\mygrad,\mygrad+120);
					\draw [gray,rounded corners,dashed,thick,fill=yellow!20](7.5,6) rectangle (14.5,11.5);
					\node at(11,11){{\large Imputing missing genes} \par};
					\node at (11,9.5) {\includegraphics[scale=1.6]{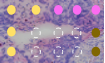}};
					\draw [dashed] (8,7.5) circle (.25cm);
					\draw [olive,fill=olive] (8,6.5) circle (.25cm);\node[right] at(8.25,6.5){{\large scRNA-seq\par}};
					\node[right] at(8.25,7.5){{\large unmeasured spots\par}};
					\node[right] at(12.5,7.5){{\large gimvI\par}};\node[right] at(12.5,6.5){{\large Tangram\par}};
					\coordinate (center) at (12,7.5);
					\fill[myred] (center) + (0, 0.25) arc (90:270:.25);
					\fill[olive] (center) + (0, -0.25) arc (270:450:0.25);
					\tkzDefPoint(12,6.5){O}
					\foreach \mycolor/\mygrad in {myred/0,myblue/90,mynaiveblue/180,olive/270}
					\tkzDrawSector[R,draw=white,fill=\mycolor](O,.25)(\mygrad,\mygrad+90);
					\draw [gray,rounded corners,dashed,thick,fill=yellow!20](15,6) rectangle (20,11.5);
					\node at(17.5,11){{\large Cell-Cell Interactions} \par};
					\node at (17.5,9.5) {\includegraphics[scale=.7]{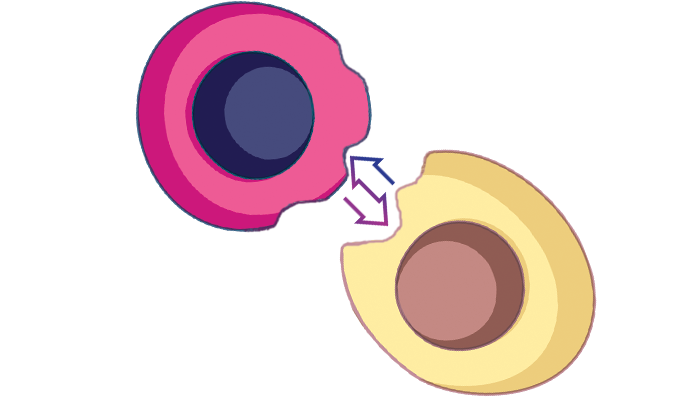}};
					\node[right] at(15,7.5){{\large stLearn\par}};
					\node [right]at(17.5,7.5){{\large GCNG \par}};
					\node [right]at(15,6.5){{\large ConST \par}};
					\tkzDefPoint(17,7.5){O}
					\foreach \mycolor/\mygrad in {myred/0,myblue/120,mynaiveblue/240}
					\tkzDrawSector[R,draw=white,fill=\mycolor](O,.25)(\mygrad,\mygrad+120);
					\coordinate (center) at (19.5,7.5);
					\fill[mynaiveblue] (center) + (0, 0.25) arc (90:270:.25);
					\fill[myred] (center) + (0, -0.25) arc (270:450:0.25);
					\tkzDefPoint(17,6.5){O}
					\foreach \mycolor/\mygrad in {myred/0,myblue/120,mynaiveblue/240}
					\tkzDrawSector[R,draw=white,fill=\mycolor](O,.25)(\mygrad,\mygrad+120);
					\draw [gray,rounded corners,dashed,thick,fill=yellow!20](20.5,6) rectangle (25,11.5);
					\node at(22.75,11){{\large Cell-type decompoition} \par};
					\tkzDefPoint(22.75,9.3){O}
					\tkzDrawSector[R,draw=white,fill=yellow](O,1.5)(340,30);
					\tkzDrawSector[R,draw=white,fill=teal](O,1.5)(30,90);
					\tkzDrawSector[R,draw=white,fill=violet](O,1.5)(90,340);
					\draw [white,fill=white] (22.75,9.3) circle [radius=0.8];
					\node at(22.75,9.3){{\large SPOT} \par};
					\node at(21.8,8.9)[rotate=-40]{{\small Cell\textsubscript{1}}};
					\node at(23.1,10.3)[rotate=-32]{{\small Cell\textsubscript{2}}};
					\node at(23.7,9.4)[rotate=-80]{\small Cell\textsubscript{3}};
					\node at(22.5,7.5){{\large GIST\par}};\node at(22.5,6.5){{\large DSTG\par}};
					\coordinate (center) at (23.5,7.5);
					\fill[myblue] (center) + (0, 0.25) arc (90:270:.25);
					\fill[myred] (center) + (0, -0.25) arc (270:450:0.25);
					\tkzDefPoint(23.5,6.5){O}
					\foreach \mycolor/\mygrad in {myred/0,olive/120,mynaiveblue/240}
					\tkzDrawSector[R,draw=white,fill=\mycolor](O,.25)(\mygrad,\mygrad+120);
					\draw [ultra thick,->](12.75,19.5)--(12.75,18.5);
				\end{tikzpicture}
			\end{adjustbox}
			\caption{Spatially resolved transcriptomics and its six sub-categories with corresponding applications. Spatially resolved transcriptomics provides gene expression profiling with spatial information in tissues. Experimentally, the measured gene expressions are captured in spots, complemented by a high-resolution histology image for the same tissue section. The resolution of spots is different (from cellular, containing multiple cells, to sub-cellular resolution, containing genes) due to the spatially resolved transcriptomic techniques. Deep learning approaches have been leveraged in spatially resolved transcriptomics data analysis to address six domains, including 1) Identifying spatial domain, 2) Identifying spatially variable genes (SVG), 3) Imputing missing genes, 4) Enhancement of gene expression resolution (GER), 5) Cell-cell interactions, 6) Cell-type decomposition. The small colored circle beside each model represents which SRT data are used in the models(blue: histology image, red: gene expression, green: spatial information).}
			\label{figure2}
		\end{figure*}

		In this section, we review 21 DL methods that have been used in the analysis of SRT data. To have a better discussion on these methods, we divided the available SRT data analysis tasks into six sub-categories; i) identifying spatial domain, ii) identifying spatially variable genes, iii) imputing missing genes, iv) enhancing gene expression resolution, v) cell-cell interactions, and vi) cell-type decomposition. A brief landscape of each sub-category and the corresponding DL are shown in Figure \ref{figure2}. To facilitate cross-references of the information, we have tabulated the utilized metrics and the dataset used in each method in Supplementary Tables 1 and 2 in supporting information, respectively, and the summary of reviewed papers in Table \ref{papers}.
		\subsection{Identifying Spatial Domain}
\begin{figure*}
			\centering
			\begin{adjustbox}{max width=\textwidth}
				\begin{tikzpicture}
					\node at (12.5,5) {\includegraphics[scale=1.2]{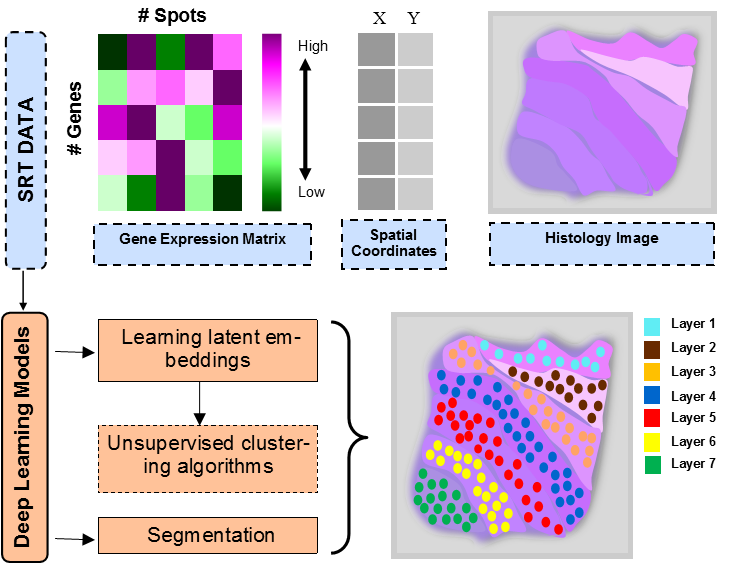}};
				\end{tikzpicture}
			\end{adjustbox}
			\caption{\textbf{Identifying spatial domains with deep learning algorithms on a synthetic tissue.} Mainly, the reviewed papers leveraged deep models to learn latent embedding and then pass them into the unsupervised clustering algorithm; or the papers leveraged deep learning models to segment the spatial domains.}
			\label{figure_cluster}
		\end{figure*}

		The spatial domain and reconstructing tissue architectures, i.e., refer to identifying spatially spots with coherent gene expression and histology, which are considered as a critical step in spatial transcriptomics analyses. Although there are many SRT platforms such as slide-seq \cite{rodriques2019slide}, which produces both tissue images and gene expression data, most SRT approaches have characterized cell types using clustering methods that use only gene expression features (i.e. Seurat \cite{stuart2019comprehensive}). Many works utilized traditional ML approaches to incorporate tissue heterogeneity and spatial information for clustering the spatial domains. \cite{zhu2018identification} developed an approach using a hidden Markov random field (HMRF) model and considered each spatial domain of cells as a set of nodes in an undirected graph and clustered each node by utilizing gene expression data. BayesSpace \cite{zhao2020bayesspace} used a Bayesian model with a Markov random field (MRF) which assumes that the gene expression data follows a multivariate normal distribution in which its latent features can be modeled by a spatial prior. Since the spatial prior does not use the exact location of spots, BayesSpace forces neighboring spots to join the same cluster. ‌Besides the high computational cost of BayesSpace and HMRF for high-resolution SRT data, MRF-based methods have the smoothness parameter, which is highly significant for determining the proximity of the neighboring spots, which is considered as a fixed value by BayesSpace. SC-MEB \cite{sc_meb} is another ML approach that is computationally efficient and adaptively learns the smoothness parameter. This method used the HMRF technique based on Empirical Bayes and utilize expectation-maximization (EM) to predict the label of each spot or cell. The above-mentioned ML approaches have an assumption on the input data, leading to knowing the prior information about the data-generating process. Thus, as we barely control the experimental design, the DL models can alternatively use because these models have minimal assumptions about the data-gathering systems. Subsequently, due to the high dimensionality of SRT data, specifically in sequencing-based methods, DL models can be effectively used because current ML methods are suitable for low-dimensional data. Thus, using DL methods for analyzing SRT data has significantly increased the functionality of finding spatial clustering. Several DL methods have been developed to account for contributing spatial data and histology images in spatial domain identification. This includes SpaCell \cite{tan2019spacell}, stLearn  \cite{pham2020stlearn}, SpaGCN \cite{hu2020integrating}, SEDR \cite{fu2021unsupervised}, STAGATE \cite{dong2021deciphering}, RESEPT \cite{chang2021define}, ECNN \cite{chelebian2021morphological}, JSTA \cite{littman2021joint}, and conST \cite{zong2022const}. Figure \ref{figure_cluster} illustrates a summary of the DL models used to identify spatial domains in SRT data. The detail about some DL models as well as their results are provided in supporting material.\\
		\subsubsection{\textbf{SpaCell.}}
		Tan et al. \cite{tan2019spacell} developed a DL model called SpaCell, which uses gene expression and tissue images. SpaCell is the first method that combines images and gene expression data in cell-type clustering. SpaCell performed pre-processing methods on the image and the count matrix separately. For histology images, it divided the entire image into small tiles, each of which is resized to $299\times299$ pixels, including one spot, then further normalization such as random rotation and Z-transform were performed. Gene counts were also mapped to each spot. For cell type clustering, the ResNet50 \cite{he2016deep} (a CNN) with pre-trained weight on the ImageNet \cite{imagenet} database was fitted on each spot to find a latent embedding vector, representing informative features in the image tile. SpaCell then imported these features, and the corresponding gene counts vector into the two AE networks and merged the obtained layers to find a latent embedding layer. K-means clustering was then applied to the embedding layer. In the disease-stage classification model, ResNet50 was fitted on the same images as the clustering step. A DNN model with two fully connected layers was then used to take the pixel features and gene count matrix as input. The output from the last layer is four probabilities, each corresponding to the four disease stages (see Supporting material for more information about the results).  However, SpaCell has two disadvantages: learning embedding without using spatial information and a pre-training model with a non-histology image which can be non-informative.
		\subsubsection{\textbf{stLearn.}}
		By inspiring SpaCell, Pham et al. \cite{pham2020stlearn} proposed stLearn, to leverage the integration of three data types optimally in SRT, including gene expression measurements, spatial distance, and tissue morphological information. The most critical aspect of stLearn is normalizing gene expression matrix via histology image. It means that stLearn tries to put the information from data heterogeneity into a gene expression matrix by normalizing it through the concept of the morphological distance of neighboring spots, in which the latent morphological features can be extracted by a CNN model (ResNet50 pre-trained on the ImageNet) and HE images as input. For each spot, stLearn identifies neighboring spots whose euclidean distance between their spatial coordinates is less than the predefined threshold. In other words, stLearn assumes that the neighboring spots (which are determined by arbitrary parameter) have similar gene expression and morphological information. This method applied global and local unsupervised clustering in two stages on the normalized gene expression data using the SMEclust function. In the former, the authors applied the PCA or uniform manifold approximation and projection (UMAP)\cite{mcinnes2018umap} methods on the normalized data followed by the KNN graph, which was constructed based on the Euclidean distance. Then, they applied k-means clustering to the graph. Regarding the latter, this work found neighbor spots with minimum distance $\epsilon=100$ for each spot and identified location-based core spots with more than $min_{samples}$. If a global cluster has more than one location-based cluster obtained by the DBSCAN algorithm \cite{schubert2017dbscan} (a clustering algorithm), it will be split into sub-clusters (see Supporting material for more information about the method and results). Like SpaCell, stLearn used an unrelated dataset for pre-train the model and also used fixed radius for identifying neighboring spots.
		\subsubsection{\textbf{SpaGCN.}}
		Hu et al. \cite{hu2020integrating} expressed that despite the efforts of stLearn in cell-type clustering using three separate input data (gene expression, spatial information, and histology image), stLearn could not link the spatial domain and biological functions. Thus, Hu et al. developed a GCN-based method called SpaGCN to identify the spatial domain by integrating the  above-mention SRT datatypes. SpaGCN constructs an undirected graph in which each node's feature is the gene expression associated with each spot. The edge's value is obtained via spatial coordinates and histological features of each spot. Unlike stLearn, SpaGCN does not limit the neighboring spot to the predefined radius and consider all spot simultaneously by weighting them in gene expression aggregation. Given the gene spatial coordinates $(x,y)$, SpaGCN added a new dimension $z$ by using the pixel coordinate and the variation of their RGB channels without using unsupervised feature learning approaches. Thus, the graph can be constructed by the Euclidean distance between the two spots, which have three dimensions. SpaGCN reduced the dimension of the gene expression matrix to 50 via PCA and utilized a GCN to link the weight of edges and gene expression for node clustering. Next, SpaGCN applied an unsupervised clustering model \cite{li2020deep}, in which each cluster represents a spatial domain containing spots with a highly correlated gene expression and histology (see Supporting material for more information about the method and results).
		Since SpaGCN used RGB channels to add an extra dimension, this approach may cause inaccuracy results due to the noisy nature of these images.
		\subsubsection{\textbf{SEDR.}}
		Fu et al. \cite{fu2021unsupervised} expressed that the SpaGCN is an oversimplified way of combining histology images with spatial information, and it needs more evidence to prove its rationality. Hence, they presented a novel method called SEDR that learns a low-dimensional latent representation of gene expression by the AE model (with two fully connected layers) and then embedded spatial data with the VGAE model (parameterized by a two-layer GCN). For each spot, the ten nearest spots consider as the neighboring spots by SEDR. The obtained embeddings from AE and VGAE were concatenated into the final latent representation; then, an unsupervised clustering method was added to obtain spatial clusters (see Supporting material for more information about the method and results). While SEDR may have justified the exclusion of histology images in their analysis, it's worth noting that both SpaCell and stLearn have demonstrated the valuable insights that can be gained from incorporating histology images, particularly in relation to tissue heterogeneity.
		\subsubsection{\textbf{STAGATE.}} 
		Despite the recent use of both data in SEDR, Dong et al. \cite{dong2021deciphering} argued that using a predefined similarity measurement between neighboring spots needs to be corrected. They proposed a graph attention autoencoder framework, STAGATE, to precisely detect spatial domains in the SRT data by incorporating spatial information and gene expression profiles. In the pre-processing step, the authors removed the area outside the tissue and used log-transformed raw gene expression as the input for STAGATE. The novelty of STAGATE mainly refers to constructing a spatial neighbor network by utilizing two approaches adaptively. The first is the standard adjacency matrix with spatial data and the predefined parameter as radius, and the second is obtained through GAT and the pre-clustered gene expression matrix. These two modules can adaptively be selected as the input of the graph attention layer. STAGATE sets the encoder into two neural network layers, where the first layer is adopted to the attention layer. Then, it performs mclust \cite{fraley2014mclust} and Louvain clustering algorithms for the labeled and unlabeled data on the learned features, respectively (see Supporting material for more information about the method and results). Despite the promising result by STAGATE, this method used a predefined radius parameter to identify neighboring spots.
		\subsubsection{\textbf{RESEPT.}}
		Although SpaGCN \cite{hu2020integrating} and stLearn \cite{pham2020stlearn} provided helpful information on spatial domain identification, Chang et al. \cite{chang2021define} argued that these methods have not revealed the intrinsic tissue architecture because of limited use of spatial information. Alternatively, they developed RESEPT \cite{chang2021define}, a DL approach for reconstructing, visualizing, and segmenting an RGB image from spatial transcriptomics. RESEPT consists of two main steps: (i) converting gene expression or RNA velocity data from spatial transcriptomics sequencing into an RGB image. This process preserves the topological relationship between spots and reconstructs the RGB image by combining spatial expression and spatial coordinates. (ii) Segmenting the image from the previous step to identify spatial domain boundaries. To achieve this, RESEPT uses a graph autoencoder, including a GCN in the encoder part, to map the input graph into a three-dimensional latent space that represents the RGB channels. The input graph is built from six neighboring spots that are adjacent in Euclidean space, using spatial information. This stage includes a backbone network (ResNet101), an encoder module (Atrous Convolutional Layers), and a decoder module. The backbone network provides rich visual features and is pre-trained using the Cityscapes dataset. The encoder module extracts multi-scale contextual information from the backbone network, while the decoder module recovers segment boundaries and spatial domains. The weights in RESEPT are optimized by minimizing the cross-entropy loss between the segmented image and the ground truth (see Supporting material for more information about the results). However, this method is limited by the fixed number of neighbors in the adjacency matrix, causing the obtained graph to be unable to learn more instructive features and biased to the model's parameters.
		\subsubsection{\textbf{ECNN.}}
		Chelebian et al. \cite{chelebian2021morphological} have investigated recent attempts in the DL area to process histology images and found a need for comprehensive methods to extract features from histology images and transcriptomics signatures. They also expressed that models need massive related data for the training process; however, current methods such as stLearn pre-trained their model on unrelated data such as ImageNet, negatively affecting the model's accuracy. Thus, they developed a modified version of the ensemble CNN \cite{strom2020artificial} previously trained on prostate needle biopsies to identify sub-regions with meaningful genetic properties in prostate histology images. As the authors in \cite{chelebian2021morphological} did not assign a name to the model developed in their work, we will subsequently refer to it as ECNN throughout this paper. ECNN used an ensemble model consisting of 30 Inception V3 \cite{szegedy2016rethinking} to classify prostate images into four classes. Chelebian et al. assumed that each model's 2048 penultimate classification layer has essential biological features that can interpret the histology images, i.e., an ensemble latent feature vector. The input images are the patches with $598\times 598$ pixels. ECNN used the UMAP feature reduction algorithm on each latent feature vector of the models ($2048\times 30$ dimension) to reduce it to $30\times 10$ descriptors per patch. This approach performed unsupervised clustering using a Gaussian Mixture Model on the ten latent features to prove that the extracted features produce relevant biological clusters to the manual annotation. Also, this work reduced the feature dimensions from ten to three, denoted as three RGB channels, to visualize the captured heterogeneity via a color map on top of the original section. A relative mean intensity (RMI) matrix was created using the measured gene expression
		factor signatures in each region to show that the obtained clusters were genetically relevant. The obtained clusters appropriately represented different gene expression factor signatures (see Supporting material for more information about the results). Although this method used a related dataset for the pre-training model, this method is blind to spatial information in SRT data.
		\subsubsection{\textbf{JSTA.}}
		Littman et al. focused on RNA hybridization-based methods, which have a high RNA capture rate \cite{lubeck2014single}. However, these methods have no prior information about the type of cell from which RNA molecules are captured. Therefore, there is a need for a comprehensive segmentation algorithm to assign the genes to the related cells. Previously, watershed-based algorithms \cite{bleau2000watershed} and the newer ML approaches were developed to segment images into cells. However, the need for experts to label the segmentations and the low quality of the data for the training step have affected these methods. Littman et al. accounted for these limitations and proposed JSTA \cite{littman2021joint}, a joint computational DL-based expectation maximization (EM) approach to enhance the segmentation of the RNA hybridization images, specifically at cellular boundaries. JSTA receives two inputs, i.e., the gene expression level of cells and pixels, described by two matrices $E_c$ and $E_p$. First, it uses the watershed algorithm on $E_p$ to obtain the initial segmentation. Then, a DNN with three layers is applied on $E_c$ to assign each gene to the related cell type with a higher likelihood. JSTA trains another DNN on the $E_p$ to obtain the cell type probability of each pixel. The two training pipelines are the E-step in the EM algorithm, which estimates the cell type distribution through pixels and gene expression. The trained pixel classifier is applied to the border pixels (determined based on the specific criteria) to re-classify those to the cell type with a higher probability. Once the JSTA re-assigned pixels in the cell borders, it updates the image segmentation and cell classifier based on the new assignments. The improvement is the M-step, which is the optimization step. The entire process is iterated until its convergence (the stable point at the end of the learning process). In both classification models, the cross-entropy loss function is used, which minimizes the error between the predicted cell type and ground truth (see Supporting material for more information about the results). However, this method lacks generalizability since it can be applied only to RNA hybridization-based methods.
		\subsubsection{\textbf{conST.}} 
		Zong et al. \cite{zong2022const} investigated the current limitations in recent papers such as SpaCell and stLearn, and proposed an interpretable multi-modal contrastive learning framework, named conST, to address three challenges in the recent SRT research, including (a) the morphological features are extracted by pre-trained CNN models (e.g., stLearn and ST-Net) or have not been involved in the training process (e.g., SEDR), (b) the biological relation between spots and the global structure of SRT data is disregarded (e.g., SpaGCN), (c) most DL models lack interpretability, preventing their further investigations in the areas that require explanation. To address these challenges, conST provides a user-friendly framework to incorporate gene expression, spatial information, and morphology information (if accessible) to learn low-dimensional embeddings for clustering and other downstream tasks. conST model the relationship between the gene expression, spatial information, and morphology within the graph and learn low-dimension representation by a general encoder $\varepsilon$. In the pre-training stage, conST used the AE model to initialize the weight of $\varepsilon$ and the dimension reduction. In the main training stage, conST used contrastive learning to learn more robust embeddings. The spots in SRT data are considered the graph's vertices, in which each node contains multiple attributes, including the gene expression matrix and the morphological feature. After performing pre-processing related to the spatial information and gene expression (see the pre-processing step in Figure \ref{figure1}), conST extracts morphological features by a masked AE (MAE) model \cite{he2021masked} pre-trained on the ImageNet dataset. conST uses the deep AE to obtain a latent embedding from gene expression.
		Meanwhile, the spatial information is passed into the VGAE to encode the position of spots into the node attribute. The latent embeddings obtained from VGAE and AE were concatenated to construct final latent features.\\ The main innovation of the conST was the use of contrastive learning in the significant step to supervise the local-local (maximizing the mutual information (MI) between similar vertices), local-global (maximizing the MI between the embeddings of each vertex and the whole graph), and local-context (maximizing the MI between the node attributes and the cluster-level summary) (see Supporting material for more information about the results). Besides using the non-histology dataset for the pre-training model, conST needs high parameter tuning, which can affect the method's functionality.
		\subsection{Identifying Spatially Variable Genes}
		\label{svg}
\begin{figure*}
			\centering
			\begin{adjustbox}{max width=\textwidth}
				\begin{tikzpicture}
					\node at (12.5,5) {\includegraphics[scale=1.2]{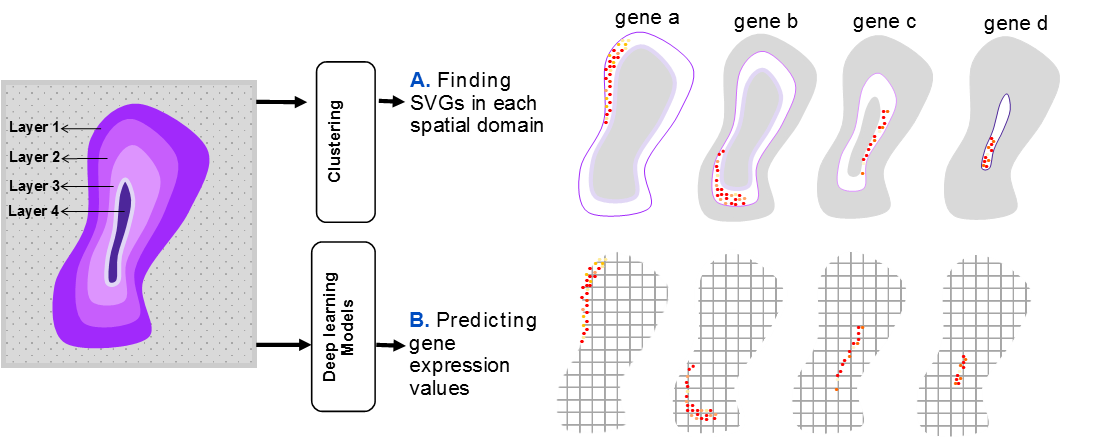}};
				\end{tikzpicture}
			\end{adjustbox}
			\caption{\textbf{Identifying spatially variable genes with deep learning methods on a synthetic tissue, including four layers (i.e. spatial domains).} \textbf{A)} Some of the reviewed papers considered this as a primary task and trained a deep model to predict the value of marker genes. \textbf{B)} Other approaches applied ML or statistical methods to identify SVGs in each spatial domain, obtained by clustering algorithms.}
			\label{figure_svg}
		\end{figure*}

		Identification and elucidating spatially variable genes (SVG) is another fundamental task in the SRT domains that aims to distinguish the spatial expression patterns across tissue sections. Recently, several statistical methods  \cite{edsgard2018identification,svensson2018spatialde,sun2020statistical} have been developed for detecting SVGs without spatial domain guidance, but they ignore tissue taxonomy. Instead, they evaluate gene associations with location independently, missing markers linked to morphological features. SVGs detection methods can be grouped into two categories based on the intrinsic principle: (i) cluster-based, which finds SVGs through statistical tests on spatial domains from clustering algorithms (e.g. SpaGCN, STAGATE, and conST); the ML part of these models are essentially trained for clustering rather than detecting SVG. (ii) whole tissue-based, which analyzes all spatially variable genes. Cluster-based methods only identify variations in discrete clusters, ignoring genes with gradient expression and samples that cannot be naturally grouped. To address this, methods that simultaneously extract features for all genes across the whole tissue are needed. ML methods in the SVG detection field can be divided into spectral and deep-based. Spectral methods aim to learn nonlinear structures by eigenvectors and eigenvalues of a positive-definite kernel.\\ RayleighSelection \cite{govek2019clustering} is a spectral-based method in SVG detection, which utilizes an extended version of the Laplacian score via a simplicial complex to calculate each gene's combinatorial Laplacian score (CLS). This method constructs the Laplacian matrix by degree and adjacency matrix (refer to GCN section). Accordingly, the genes with the lowest CLS are more spatially variable. RayleighSelection was tested on the FISH dataset, and genes with high consistency to the spatial expression pattern were obtained. Nevertheless, spectral-based methods are computationally expensive and must scale up for large datasets.
		Since the H\&E-stained histology images are cheaper and more accessible than spatial transcriptomics data, it makes them desirable to jointly leverage with SRT data. Fluorescence in-situ hybridization and in-situ sequencing techniques complement spatial transcriptomics, which captures more genes with low spatial resolution along with tissue images. Specifically, (ISH) captures high-resolution images of spatial gene expression at cellular resolution.  Due to the complex nature of these images and the need for the full exploitation, deep learning based methods have performed better than spectral technique-based methods.
		Concerning these limitations and the influence of image histology, DL methods have been proposed for SVG detection both in cluster-based domain, such as CoSTA and deep-based domain, including ST-NET, SPADE, HisToGene, CNNTL, and DeepSpaCE. Figure \ref{figure_svg} shows the overall view of distinguishing SVGs with DL models, reviewed in this paper.  
		\subsubsection{\textbf{SpaGCN.}}
		Hu et al. \cite{hu2020integrating} proposed SpaGCN to detect SVGs in each obtained cluster (refer to the previous section). As the second pipeline, SpaGCN identifies SVGs which are highly expressed in each spatial domain. Considering the spatial domains obtained by clustering, SpaGCN first finds the neighboring domains (clusters) of a targeted domain based on a defined criterion (e.g., distance) and selected SVGs based on the Wilcoxon rank-sum test. Additionally, the authors noted that some genes may be expressed in multiple but scattered domains and called those metagenes, which are still helpful for understanding the spatial variation of gene expression. Briefly, SpaGCN first identifies genes with a weaker expression level than SVGs in the target domain by reducing threshold values in the previous pipeline. SpaGCN then randomly selects the genes with a corresponding mean value of gene expression as the base genes and identifies the genes with higher expression values in the target and not-target domains as positive and negative genes, respectively. SpaGCN next adds the positive genes to and subtracts the negative genes from the base genes to detect the metagenes uniquely expressed in the target domain. The entire process would iterate for all genes in the target domain. Moreover, SpaGCN provides the sub-cluster option to the obtained spatial domain by using the information from neighboring spots to characterize heterogeneity across the spatial domain. The performance of SpaGCN in SVG detection is measured by Moran’s I \cite{li2007beyond} statistic metric and SpaGCN achieved more coherent SVGs with better biological interpretability than the SPARK \cite{sun2020statistical} and SpatialDE (statistical methods). SpaGCN identifies SVGs as a downstream task, and the deep model is essentially trained to clustering rather than design for SVG detection. Thus, the obtained marker genes do not associate with tissue heterogeneity.
		\subsubsection{\textbf{conST.}}
		Zong et al. \cite{zong2022const} detected spatial marker genes on the obtained clusters (refer to the previous section) and considered it as a downstream task to show the accuracy of the clustering results. conST uses a similar approach like SpaGCN into the latent embedding obtained from the main algorithm. (refer to the conST method in the spatial domain identification section). The authors compared conST with obtained marker genes by SpaGCN in terms of Moran’s I on the spatialLIBD \cite{maynard2021transcriptome} dataset, in which conST achieved better performance, particularly at the boundary of white matter layer.
		\subsubsection{\textbf{STAGATE.}}
		STAGATE is another deep model which identifies SVG in the spatial domain, however, it has not been trained for this task. Similar to SpaGCN, obtained SVGs do not correlate with morphology. STAGATE applied the Wilcoxon test implemented in the SCANPY package to identify spatially variable genes for each spatial domain. STAGATE could also detect SVGs in the Slide-seqV2 dataset from mouse olfactory bulb tissue, and it can detect more genes in the small tissue structures than the SPARK-X algorithm.
		\subsubsection{\textbf{CoSTA.}}
		CoSTA is a cluster-based method that avoids using hierarchical clustering and treats pixels as independent features. It uses an expression matrix constructed for gene expression analysis and employs PCA as a pre-processing step, which takes into account pixel-level correlations instead of preserving the spatial relationships between neighboring cells. \cite{eng2019transcriptome}. 
		Xu et al. \cite{xu2021costa} claimed that most spatial patterns become lost by these approaches. They proposed CoSTA that uses an unsupervised convolutional neural network learning approach to learn spatial relationships between genes by using more information about the positions of neighboring pixels in spatial transcriptomic images. In the pre-processing step, 100 pixels are binned into one pixel and resized into the $48 \times 48$ image size. After binning, CoSTA normalizes gene matrices as described in \cite{svensson2018spatialde} and scales them between 0 and 1 through dividing the gene matrices by the maximum value of the $48 \times 48$ matrix. CoSTA consists of two steps: clustering and neural network training. In the first step, CoSTA passes the normalized images through the CNN network (ConvNet), which consists of three convolution boxes (each box contains Convolution, batch normalization and max-pooling layer). To cluster features, the method applies the L2-normalization and UMAP (an unsupervised dimension reduction method) to the feature vector, respectively. The purpose of clustering is to generate labels for training ConvNet. Once the label generation in the first step is finished, the second step adds a fully connected layer (FC), with the Softmax activation function, to the last layer to produces the probability of the input gene belonging to each cluster. The method uses the FC layer just during training, and it would be discarded in the first mentioned step. CoSTA recognizes the quantitative similarity between genes in the MERFISH dataset, which achieves better results than SpatialDE and Spark. The CoSTA results on Slide‑seq data demonstrated that it identified spatial patterns‑dependent, accurately (see Supporting material for more information about the method and results). However, this method needs high parameter tuning and only was assessed on high-resolution SRT data. 
		\subsubsection{\textbf{ST-Net.}}
		He et al. \cite{he2020integrating} proposed a deep model that exploits whole tissue called ST-Net to integrate spatial transcriptomics and histology images in predicting high-resolution gene expression in patients with breast cancer. To account for preparing images in the training step, ST-Net tiles the input image with $10000\times 10000$ pixels into the $224\times 224$ pixels centered on the spots and calls them patches. Next, ST-Net adds a pseudo count of one to the gene count from each spot and normalizes it within the log transformation. The normalized patches across the whole slide are the network's input. After normalization, ST-Net trains the CNN network consisting of DenseNet-121 with pre-trained ImageNet weights followed by two fully connected layers with 1024 and 250 vector lengths representing the latent feature vector and the predicted gene expression from 250 genes, respectively. Mainly, ST-Net treats this problem as a multivariate regression problem. 
		The authors trained ST-Net on a breast cancer spatial transcriptomics dataset and achieved a mean square error of 0.31 and a Pearson's correlation of 0.33 (the average of all 234 genes), with 102 of the 250 genes positively correlated in nearly 20 patients. The UMAP visualization then showed that the latent feature vectors can distinguish between tumor and non-tumor spots, which has potential applications in clustering and cell-type composition (see Supporting material for more information about the results). Despite its robustness and generalizability, this study does not fully exploit the spatial information available in the data.
		
		\subsubsection{\textbf{SPADE.}}
		The combination of image and spatial gene expression data have provided complementary information about morphological patterns in tissue. Bae et al. \cite{bae2021discovery} used morphological heterogeneity to detect SVGs in which each gene was considered a dependent variable in the model. They proposed SPADE, a convolutional neural network model, to extract features from image patches around each spot and combine them with gene expression data to obtain spatial marker genes. In contrast with ST-NET that utilizes tissue images to predict marker genes, SPADE exploits the relationship between morphology and gene features. In the first step, SPADE extracts features from patches by utilizing VGG-16, in which the weights were pre-trained on the ImageNET dataset. The last layer of VGG is a 512-dimensional vector, which the dimension is reduced by the PCA algorithm. The number of principal components (PCs) varies regarding the input dataset. After normalizing genes in each spot, the authors used Limma \cite{ritchie2015limma} for discovering SVGs, and a linear regression model to fit normalized genes to PCs of latent image features. The linear model's goal is to rank the associated genes base on the PC's value regarding regression coefficient (RC) or corrected P-value. Ultimately, the genes are selected as the spatial maker genes that present a false discovery rate (FDR) less than 0.05 in PCs which explains more than 2\% of the variance in 512-dimensional image features. SPADE identifies marker genes positively associated with the endoplasmic reticulum, synapse organization, and cell adhesion molecule binding in human breast cancer, olfactory bulb, and prostate cancer dataset, respectively. Nonetheless, The obtained marker genes can be significantly affected by the spot's density and the distance between spots.
		
		\subsubsection{\textbf{HisToGene.}}
		\label{histo}
		ST-Net has shown that tumour-related genes are highly correlated with the histology images, however this method did not use spatial location information in their CNN-based model. Pang et al. \cite{pang2021leveraging} declared that despite the CNN performance in the image processing tasks, it suffers from intrinsic bias regarding the SRT patch position. Thus, this drawback reduces the CNN model's effectiveness on SRT data. Consequently, they used an autoencoder model with an attention-based mechanism \cite{vaswani2017attention} called HisToGene \cite{pang2021leveraging} to predict gene expression values by embedding spatial location and histology images. The pre-processing step comprised of the following steps: removing genes with low expression that appeared in fewer than 1000 spots, normalizing the UMI count of each gene by dividing it by the total UMI count across all genes in the spot and multiplying by 1,000,000, and transforming to a natural log scale. The HisToGene method extracted patches from the histology images and transformed the images into a new matrix. The authors acknowledged that, like in natural language processing (NLP) where sentences of varying length exist, the number of spots within a tissue also varies, making it unsuitable to split them into a fixed number of patches. Thus, they modified the encoder part, encoding the new image matrix and spatial coordinate through a single layer encoder. The final embedding matrix was obtained by summing up the encoding matrixes, considered an input for the multi-head attention layers (see \cite{vaswani2017attention} for more details about attention models). The multi-head attention module consists of eight multi-head attention layers and 16 attention heads, automatically learning the gene expression from sequencing patches (see Supporting material for more information about the results). The authors compared the HisToGene with ST-Net, which consistently outperformed ST-Net in correlation, but was assessed only on high-resolution SRT data.
		\subsubsection{\textbf{CNNTL.}}
		Abed-Esfahani et al. \cite {pegah} noted that as each image in the ISH method represents a specific gene, and there are only a limited number of images per gene, alternative methods to classification-based approaches may be more suitable. Therefore, they proposed a CNN-based method using contrastive loss to re-identify gene expression and embed gene expression patterns from the human brain. As they did not choose a name for the proposed model, we named the proposed model CNNTL (CNN with triplet loss) to better refer to the model in this study. In the pre-processing step, CNNTL imported the U-Net model with ResNet34 as a backbone to separate grey and white matter from the background in images. The segmented image was tiled into patches to contain at least 90\% of the foreground. In the training step, CNNTL leveraged the triplet loss. The loss function ensures that the learned embedded from an input patch (positive image) is closer to another patch from the same class (anchor image) compared to the patch that belongs to another class (negative image). The learned embedded of three input images were obtained from three ResNet50 models (with shared weights), pre-trained on the ImageNet \cite{imagenet}, followed by two fully connected layers with the dimensions of 1024 and 128, respectively. The CNNTL approach was tested on the Cortex dataset obtained from 42 donors. The metric is rank-1 accuracy at the level of images, which is the proportion of images for which the Euclidean distance of their embeddings computes the closest image of the same gene. The CNNTL achieved rank-1 accuracy of 38.3\%, which performed better than single ResNet or random models (see Supporting material for more information about the results). Despite the novelty of CNNTl, The method is limited to a small fraction of genes in brain layers and pre-trained on a dataset unrelated to SRT data.
		\subsubsection{\textbf{DeepSpaCE,}}
		\label{deepspa}
		Monjo et al. \cite{deepspace} focused on the in situ capturing technology in SRT due to this method's significant effect in oncology. They developed a convolutional neural network model named DeepSpaCE to predict gene expression.
		DeepSpaCE is a VGG16 network that takes spot images as an input and predicts the expression of 24 genes, including breast cancer-marker genes and breast cancer-related micro-environment marker genes. The DeepSpaCE was tested on a dataset from human breast cancer and obtained 0.588 correlation coefficients between the measured and predicted values (see Supporting material for more information about the results). However, the model has limitations in that it can only predict a limited number of genes.
		\subsection{Imputing Missing Genes}
\begin{figure*}
			\centering
			\begin{adjustbox}{max width=\textwidth}
				\begin{tikzpicture}
					\node at (12.5,5) {\includegraphics[scale=1.2]{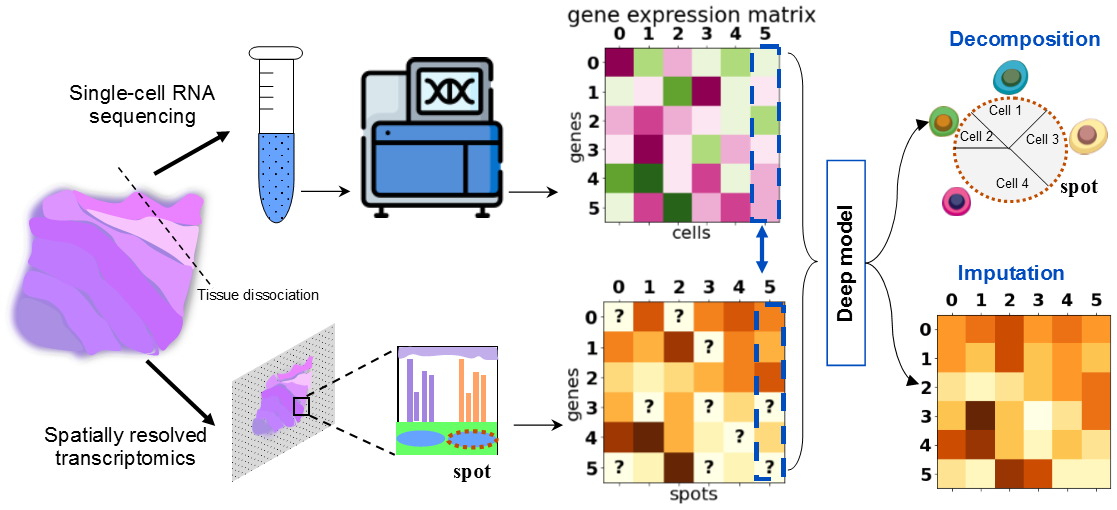}};
				\end{tikzpicture}
			\end{adjustbox}
			\caption{\textbf{Imputing missing genes and cell-type decomposition with deep learning models on a synthetic tissue.} In the sequencing-based approaches, the transcriptomes are captured on the spots, which cannot measure all genes inside a tissue. On the other hand, single-cell sequencing can perform this task with tissue dissociation, leading to spatial information loss. Thus, using the SRT and scRNA-seq data from the same tissue, the deep learning models can measure the missing genes and the proportion of cells in each spot.}
			\label{figure_45}
		\end{figure*}

		Single-cell RNA sequencing (scRNA-seq) is a sophisticated technique that provides an unrivaled expression profile of a considerable number of genes across tissues at the resolution of an individual cell \cite{single_cell}. However, a drawback of these methods is the requisite sample distinction, which destroys any spatial context, which can be crucial to understanding cellular attributes \cite{yuan2017challenges}. In contrast, SRT data can capture cell location but is limited to the resolution of SRT technology. Recent papers have proposed computational approaches to integrate scRNA-seq data and spatial transcriptomics to predict unmeasured genes and impute gene expression in spatial data. Many machine learning methods employ joint dimension reduction and joint embedding projection to integrate the scRNA-seq and spatial transcriptomics data, followed by using the K-Nearest Neighbor (K-NN) algorithm to predict the missing (unmeasured) genes in spatial transcriptomics data.
		LIGER \cite{welch2019single}, and SpaGE \cite{abdelaal2020spage} are two ML methods that utilized joint dimension reduction approaches, including NMF and PCA, respectively, then learn joint embedding by linear models. Lopez et al.  \cite{lopez2019joint} demonstrated that a portion of genes can be found in both scRNA-seq and spatial transcriptomics data, and therefore, domain adaptation methods can be a solution. They proposed gimVI, a joint non-linear model based on deep generative models. Shengquan et al. \cite{shengquan2021stplus} critiqued recent approaches that only utilized genes shared between both datasets and employed unsuitable evaluation metrics, such as the Spearman correlation coefficient. This metric can be misleading in indicating the actual performance of a method, as even though the Spearman correlation coefficients may be low, visual exploration may reveal improved patterns for known genes \cite{lopez2019joint}; thus, they developed an AE model called stPlus \cite{shengquan2021stplus} to predict gene expression using the learned embedding and k-NN algorithm.
		stPlus applied Louvain clustering on the predicted genes and compared it with current ML methods based on the four clustering metrics AMI, ARI, Homo, and NMI (see Supplementary Table 1). The proposed method showed better results than SpaGE, Seurat, Liger, and gimVI in all four metrics.\\
		In the following, we focus on the DL models and investigate the three DL models in detail. Figure \ref{figure_45} represents the process of gene imputation along with cell type decomposition (refer to the next section) in SRT data.
		\subsubsection{\textbf{gimVI.}}
		gimVI specifies the binary variable $s_n$ for each cell, denoted as whether the scRNA-seq or the SRT experiment captured the cell, and models the gene expression matrix with either NB or zero-inflated NB (ZINB). The generative model is a VAE, which gets the input cells and $s_n$. The output from the encoder part is a latent vector $z_n$, describing the cell type $n$. gimVI then measures the probability of each gene $g$ in an individual cell from the decoder part. Finally, they impute missing genes by implementing the K-NN algorithm on obtained latent space. Lopez et al. evaluated the gimVI on the two paired datasets of scRNA-seq/SRT. They calculated the spearman correlation to assess the performance of the gene imputation process in 20\% of the genes in the SRT dataset. The results showed that gimVI works much more efficiently in imputation than the Liger and Seurat. However, the reported results can vary by the number of K and evaluate the model on the small fraction of genes.
		\subsubsection{\textbf{Tangram.}}
		\cite{biancalani2021deep} developed the Tangram model to map spatial information into the scRNA-seq data and align the histological data to the anatomical position via a DL framework. Technically, Tangram is a DL tool for aligning sc/snRNA-seq data to spatial data by utilizing nonconvex optimization. The input of Tangram is sc/
		snRNA-seq data and SRT data from the same region or tissue type, and the output is a matrix that contains the probability of assigning each cell in sc/snRNA-seq data to the voxel of SRT data. Tangram first randomly mapp sc/snRNA-seq data to the space. Then their alignment is updated through an objective function. In the nonconvex optimization process, Tangram aims to compare cell-density distributions of sc/snRNA-seq and SRT data using KL divergence, whereas gene expression is assessed by cosine similarity. Although Tangram was specifically developed for the reconstruction of spatial maps, the imputation task was effectively performed in the intermediate process step. Therefore, the original paper has no quantitative comparison between Tangram and other imputation methods. \cite{impute2023} recently prepared a comparative performance evaluation for imputation methods, and Tangram has placed third-best method after stPlus and gimVI. However, it has been shown that Tangram has the highest running time compared to the other imputation methods. \\
		
		\subsection{Cell-type Decomposition}
		In the spatial transcriptomics method, transcripts are captured at spatial locations, called spots \cite{staahl2016visualization}, and often consist of a mixture of low-resolution cells (such as sequencing-based and ST/Visium technologies). The number of cells is different due to the tissue heterogeneity or SRT technology \cite{saiselet2020transcriptional}. Therefore, it is important to identify the cell composition in SRT data at the spot level (Figure \ref{figure_45}). Recently, various computational methods have been developed for this purpose, grouped into three main categories: 1) Inference-based methods, 2) Multivariate analysis and linear algebra-based methods such as SPOTlight \cite{elosua2021spotlight} and SpatialDWLS \cite{dong2021spatialdwls}, and 3) Deep learning-based methods. Inference-based methods, including Stereoscope \cite{andersson2020single}, RCTD \cite{cable2022robust}, cell2location \cite{kleshchevnikov2022cell2location}, DestVI \cite{lopez2022destvi}, and STdeconvolve \cite{miller2022reference}, utilize likelihood-based approaches and explicitly or parametrically assume the data distribution in the input data. These methods belong to both machine learning and statistical approaches, with limitations discussed in Section \ref{sec1}. Meanwhile, deep-based methods such as GIST, DSTG, and Tangram, estimate cell-type proportions using deep learning models. Some deep learning methods, such as VAE, are also based on probabilities but are still considered deep-based methods in this paper. However, these deep learning methods have limitations in real-world applications.
		
		\subsubsection{\textbf{DSTG.}} 
		
		The DSTG method, proposed by Song et al. \cite{DSTG}, is a graph convolution network that uses a semi-supervised approach to decompose cell mixtures in SRT (spatial transcriptomics) data. It creates pseudo-ST data from scRNAseq data, projects the data into a 20-dimensional space, and builds a linked graph. The linked graph, represented as an adjacency matrix $A$, and the data matrix $X$ are fed into a GCN network consisting of three convolution layers. The output of the network is the predicted proportions of different cell types in the pseudo and real SRT data, which are learned by minimizing cross-entropy loss with the ground truth. The evaluation results show that DSTG outperforms the SPOTlight method on both synthetic and real SRT datasets. The evaluation process shows that DSTG obtains better results than SPOTlight on synthetic and real SRT datasets. However, utilizing the Euclidean distance to show the similarity between the pseudo-ST and real-ST is not a fair comparison.
		\subsubsection{\textbf{Tangram.}}
		In Tangram, the authors demonstrated that inference-based deconvolution methods can be limited by the lack of use of spatial information, resulting in inaccurate detection of cell types defined by sparse markers. Tangram performs deconvolution on ST/Visium technology, considered a low-resolution SRT method. Tangram first calculates the number of cells by performing initial segmentation, and then passes the segmentation results to the Tangram model (see the previous section) to calculate the cell fraction per spot. The results on three visium datasets showed that Tangram was able to find consistent mapped cell-type ratios and those from the snRNA-seq data. However, Tangram required pre-knowledge about the cell numbers to perform segmentation before deconvolution, which may not be easily obtained in higher-density tissues, such as tumors \cite{biancalani2021deep}.
		\subsubsection{\textbf{GIST.}} 
		In the study by Zubair et al. \cite{zubair2021jointly}, a joint model was presented to improve cell-type decomposition by integrating gene expression data from spatial transcriptomics (SRT) and image-derived data from the same tissue. The main objective of the model, named Guiding Image-based ST (GIST), was to leverage deep learning (DL) on images as preliminary information in a Bayesian probabilistic model for cell type identification. GIST utilized a DL approach to estimate the abundance of cell type A (e.g. immune cells) at a given spot by using a convolutional neural network (CNN) model. The JPEG format of the images was first converted to an encoded tiled TIFF format and then fed to a pre-trained VGG16 model on the TCGA dataset. This generated the probability of the patches of 50x50 microns. The spot-level probability was then obtained by a weighted sum of overlapping patches across the spot. GIST used the estimated cell-type proportion from DL as an informative prior distribution and mapped it onto the first round of model fitting distribution in the GIST base-model, which was trained with SRT data without prior information. The results showed that the GIST model improved the identification of immune cells in pathologist-annotated regions compared to the GIST base model using expression data only. The performance of the GIST model was demonstrated in breast cancer pathology, and it may be generalizable to immunofluorescence. However, a comparison to current deep-based methods, such as Tangram and DSTG, is needed.
		
		\subsection{Enhancement of Gene Expression Resolution}
\begin{figure*}
			\centering
			\begin{adjustbox}{max width=\textwidth}
				\begin{tikzpicture}
					\node at (12.5,5) {\includegraphics[scale=1.2]{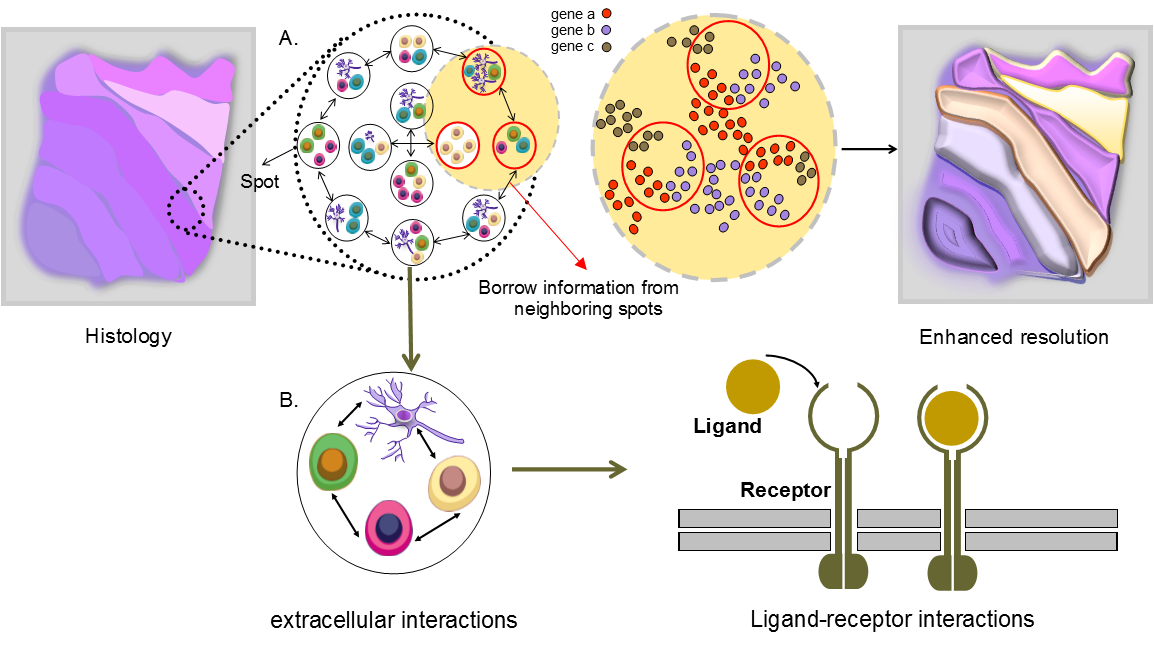}};
				\end{tikzpicture}
			\end{adjustbox}
			\caption{\textbf{Enhancing gene expression resolution and cell-cell interactions in SRT data. A)} Since the distances between spots are different based on the utilized sequencing-based approaches, borrowing information from neighboring spots makes it possible to enhance the gene expression resolution in empty areas between spots. \textbf{B)} The spatial location of each spot facilitates the understanding of finding ligand-receptor interactions of each cell in SRT data.}
			\label{figure_eoger}
		\end{figure*}

		Sequencing-based spatially resolved transcriptomics data often have limited resolution at the single-cell level. To improve gene expression resolution in SRT data, various deep learning methods have been proposed to borrow information from neighboring areas to fill the gaps between spots and enhance gene expression resolution. The overall process of this approach, including cell-cell interactions, is depicted in Figure \ref{figure_eoger}. High-resolution information on cell morphology is available in some popular SRT technologies, such as Visium and SLIDE-seq \cite{rodriques2019slide}, as histology images from H\&E-stained tissue sections. However, statistical methods like RCTD that estimate cell-type-specific gene expression for each spot based on the probability of cells obtained in the deconvolution process are unreliable, as the results depend on the accuracy of the deconvolution step. BayesSpace resolves this issue by dividing each spot into multiple equal-size sub-spots and inferring gene expression while keeping the total expression of the original spot constant. However, different splitting methods may produce different results, making it challenging to determine the optimal solution. Due to the capability of DL methods to integrate multiple data, the following DL methods have utilized histology images in enhancing the gene expression resolution, in which none of the above methods take advantage of this.
		
		\subsubsection{\textbf{XFuse.}}
		Bergenstrahle et al. \cite{xfuse} developed XFuse to infer high-resolution spatial gene expression from the histology image data by integrating low-resolution gene expression from in situ sequencing and high-resolution histology images. XFuse assumes that the conditional distribution of gene expression data and histology images ($I$) follow negative binomial (NB) and gaussian distribution, respectively. Next, it maps the parameters of the mentioned distributions from the latent tissue state ($Z$) via a convolutional generator network $G$. Then, XFuse uses variational inference to estimate the posterior of the latent variable ($N(Z \mid \bar{X}, I)$), where $\bar{X}$ is the observed expression data at a specific location. It updates the variational and network parameters by minimizing the Kullback-Leibler divergence, which measures the differences between two distributions. Simultaneously, XFuse amortises the inference through a convolutional recognition network on histology images to map them to the latent tissue state. The authors assessed the performance of the XFuse model on mouse olfactory bulb and human breast cancer datasets. The results revealed that XFuse was able to uncover distinct patterns in both datasets, outperforming a method that used non-missing neighbors' information to fill in missing data. XFuse was also found to have a lower median root-mean-square error (RMSE) and to accurately predict unseen samples. When compared to in situ hybridization data, XFuse showed better prediction of gene expression patterns in the tissue. However, the model has a limitation in that it can only detect genes whose spatial patterns are similar to the histology images.
		
		\subsubsection{\textbf{HisToGene*.}}
		The resulting super-resolution prediction pipeline in HisToGene (see identifying SVGs section) was named HisToGene*. The HisToGene* study presents a novel approach for super-resolution gene expression prediction by averaging the predicted gene expression from dense histology image patches. The authors applied the trained HisToGene model to images and estimated gene expression at the spot-level resolution. They then treated spots as sentences in NLP and created sub-patches covering four patches each to predict gene expression at a higher resolution than the original spot. The experiments from the HisToGene study were repeated in the HisToGene* study. The results showed that the HisToGene* predicted spot-level gene expression had higher correlations with the observed spot-level gene expression in 19 sections compared to HisToGene. In the remaining six sections, HisToGene showed higher correlations than HisToGene*. The study also observed a direct link between thyroid hormones and the risk of breast cancer in the HisToGene* top enriched pathways \cite{ortega2018thyroid}, indicating that the predicted gene sets by HisToGene* contain more biologically meaningful information.
		
		\subsubsection{\textbf{DeepSpaCE.}}
		DeepSpaCE is a method that utilizes super-resolution of spatial gene expression and imputation of tissue sections to predict gene expression. It uses a trained model to estimate unmeasured genes in images with insufficient gene expression. The method leverages semi-supervised learning (SSL) to improve its performance. The DeepSpaCE method was tested on a human breast cancer dataset that consisted of three tissue sections (A, B, C) and related consecutive sections (D1-D3). The model was trained on sections C and D2 for the super-resolution step and sections D1 and D3 were used as a training set for the section imputation step, with section D2 as the test set. In this case, the model acted as a "teacher" model in SSL. Using sections A, B, and C as unlabeled data, the Pearson's correlation coefficients (PCC) between actual and predicted expression were increased. The SSL approach was also applied to cat and dog images from ImageNet and the Cancer Genome Atlas (TCGA) dataset, but no improvement was observed in the results.
		
		\subsection{Cell-Cell Interactions}
		Cell-cell interaction refers to the communication between cells through the binding of a ligand to its complementary receptor, a process known as ligand-receptor interaction (LRI). LRIs are crucial for extracellular communication and have been studied as a way to understand cell-cell interactions  \cite{GURYANOV2016890}. However, most current computational methods in this area either focus on intracellular interactions or are limited to investigation of small-scale experiments. Spatial transcriptomics provides gene expression profiles in spatial coordinates within individual cells, making it a potentially valuable tool for predicting LRIs (see Figure \ref{figure_eoger}). The Giotto \cite{dries2021giotto} toolbox is a comprehensive framework for analyzing spatial transcriptomics data, including a module for cell-cell interactions. Giotto and other statistical methods identify interactions within a cellular niche by constructing a model for the expression of markers. MISTy \cite{misty} is a scalable machine learning framework that can identify a range of cell-cell interactions in spatial transcriptomics data by generating pairwise distances. However, it is computationally extensive. The unique feature of MISTy is identifying CCI within specific regions of interest that facilitate understanding of the marker interactions, which have not been considered by the deep learning based methods. Additionally, the performance of non-DL models is affected by the growth of SRT data from different species and tissue in size and resolution. At this point, the DL methods can provide a better solution to identify cell interactions in the large SRT data.
		\subsubsection{\textbf{stLearn.}}
		stLearn presents a method to analyze cell type diversity and RLIs separately and then combine them into an interaction measured through a whole tissue section. The algorithm has two steps: i) quantifying cell type diversity by dividing the tissue into windows and counting the cell types of interest, and ii) finding RLIs by calculating the co-expression of ligand and receptor pairs in the central spot and its neighbors using the CellPhoneDB algorithm \cite{efremova2020cellphonedb}. Given the expression threshold, the co-expressing L-R pairs for the central spot are calculated by Eq(see supporting material). A CCI matrix is then generates to show the significant L-R pairs for each spot, and clustering performs to identify tissue regions based on the most similar L-R co-expression values.
		Finally, incorporating both cell density and CCI measures, stLearn can identify tissue regions with a high likelihood of cell-cell interaction. The performance of the CCI algorithm was tested on a breast cancer dataset and was able to detect high interaction between tumour and immune cells. \
		
		\subsubsection{\textbf{GCNG.}}
		Yuan and Bar-Joseph \cite{yuan2020gcng} reported that the recent models, such as Giotto, have mainly concentrated on unsupervised correlation-based analysis in detecting extracellular interactions, leading to failure in predicting interactions in a complex pattern. To address this, they proposed a graph convolutional network for gene expression (GCNG) model, which predicts extracellular interactions from Spatial Transcriptomics data. GCNG is a five-layer graph convolutional network consisting of two CNN layers, a flattening layer, and a sigmoid activation function layer to calculate the probability of ligand-receptor interactions within cells. The network takes two inputs, including the spatial location of cells (neighborhood graph) and gene expression pairs in each cell. First, it constructs an adjacency matrix $A^{R\times R}$ from the total cell number $R$ in which the element is $1$ if the Euclidean distance of two cells in the spatial location is smaller than the predefined distance threshold and $0$. Second, the input matrix $X^{R\times 2}$ is built based on paired candidate genes in each cell. The two matrices are multiplied by each other in the first layer and then mapped to the embedding vector, leading to the investigation of more interactions between cells without a direct link. Finally, the output predicts the probability of interactions between two paired genes. The model tested on the various normalization types of matrix $A$, and the GCNG reached median AUROC/AUPRC of 0.99/1.0 and 0.99/1.0 for seqFISH+ and MERFISH, respectively. The model outperformed the recent intercellular communication models such as Giotto or the single-cell Pearson correlation between ligand and receptors method. However, GCNG selects predefined distance criteria for selecting the neighbor cells, which may cause biases in the obtained results.
		
		\subsubsection{\textbf{conST.}}
		conST leverages the advantage of clustering, SVG detection, and trajectory inference for identification of target receptors on breast cancer cells and analysis of their microenvironment in IDC regions. To do this, conST first obtains latent features from the breast cancer dataset and clusters them into 20 clusters. Then, it detects three clusters containing the prominent lesion areas and applies trajectory inference to obtain pseudotime ordering. The SVG detection algorithm (see conST in the SVG detection section) is applied to detect marker genes responsible for the tumour microenvironment. Finally, cross-cluster CCI analysis is performed using TraSig \cite{weiler2021guide} and within-cluster analysis is done by label transfer from Seurat to detect active ligand-receptor pairs. The results demonstrated that conST can successfully detect IDC, DCIS, and edge tumour cell regions in cross-cluster analysis and active L-R pairs in within IDC regions Within-cluster analysis.
		
		\section{Discussion and Future Look }
		
		We outlined the advantages and disadvantages of available DL algorithms for analyzing imaging- and sequencing-based SRT data (Table \ref{papers}). Additionally, we provided a comprehensive technical overview to understand the performance of each method (see Supplementary Table 1). We further contrasted non-DL and DL methods to emphasize how DL methods can improve the analysis process of SRT data. As many downstream tasks rely on the individual components derived from the entire workflow, the downstream analyses will be negatively affected if a component does not work optimally. For example, the identification of SVGs is reliant on the clustering algorithm. The downstream analysis will be impacted if the clustering algorithm has neglected a biologically relevant feature. Also, phylogeny-aware clustering is increasingly used for biological data sets by incorporating the phylogeny of the organisms into their effect size on creating the resemblance matrix. Therefore, we suggest using phylogeny-aware clustering techniques by incorporating "pathway information" into SRT data \cite{lotfollahi2023biologically}. One way to do this is by incorporating KEGG-level pathways or other reference assignments into the effect size of a gene. If two genes produce proteins that function within the same pathway, then both of them changing is less impactful to the overall picture than two genes from completely different pathways.\\ Also the reviewed techniques could not leverage the full advantages of rich information in the SRT data. Therefore, there is still a need for more robust DL methods to jointly use spatial data, scRNA-seq, and high-resolution histology image data. Most of the reviewed techniques in this study were developed based on histology images. As the CNN networks have mainly obtained promising results in image processing, these networks were more substantially used in the SRT data analysis than the other deep models. However, histology images' unique characteristics and complex structures (e.g., irregularity and large scale) pose challenges in DL algorithms, particularly when integrating with spatial data.
		We believe there is a substantial need to prove that the extracted features by deep models are biologically meaningful. For example, ECNN \cite{chelebian2021morphological} used the CNN intermediate layers to plot the features, or ST-NET \cite{he2020integrating} also used features from the latent vector to plot the 2D UMAP. Additionally, due to the large histology images, all methods in this review tiled the input images into small patches. Effective analysis of the histology images requires new methods to treat the extracted patches as united data, i.e., by treating them as a word used in the NLP techniques \cite{pang2021leveraging}. Although HisToGene \cite{pang2021leveraging} can link the patches by developing an attention-based model, other techniques could not consider the relation between patches through their deep algorithms. This problem is most important in CNN-based models, making them more sensitive to the influence of the batch effect \cite{pang2021leveraging}. Thus, it will be desirable to consider patches as time-series problems and develop DL methods on sequential data such as recurrent neural network (RNN), long short-term memory (LSTM) \cite{hochreiter1997long}, and transformers.\\
		Another essential issue in SRT data processing is the batch effect, which becomes more apparent due to the abundance of spatial transcriptomic datasets. Although several DL methods have been developed for batch effect correction in scRNA-seq \cite{li2020deep,lakkis2021joint}, the SRT domain still suffers from generating a deep model to address the batch effect challenges. Importantly, this problem is even more complex in SRT data because of spatial dependency and histology images association, in which batch effects can affect both gene expression and histology images. SEDR \cite{fu2021unsupervised} was the first deep model for batch effect correction, which uses the SEDR-derived embedding and the Harmony algorithm \cite{tran2020benchmark} for batch effect removal. STAGATE \cite{dong2021deciphering} was another attempt to extract the 3D expression domain and reduce the batch effect between consecutive tissue sections. However, neither of the two mentioned algorithms accounts for histology images. Since the batch effect may also occur on associate histology images, methods to jointly evaluate gene expression and histology images are required to alleviate the batch effect across the tissue sections as well as between them, simultaneously.\\
		Histopathology has provided a comprehensive perspective across various medicine domains such as disease staging and cancer development in tissues. As such, it is named the gold standard in diagnosing almost all types of cancers \cite{washington2010ajcc}. Interestingly, Spatially resolved transcriptomics has enabled the analysis of both imaging and molecular features. SpaCell was the first DL model for cancer stage classification using both image and gene expression. CNNTL \cite{pegah} was another deep model that classified the Schizophrenia patient into control and non-control by using image-based SRT data. With the development of SRT technology and reducing the cost of SRT data-generating, it would be cutting-edge technology to diagnose diseases routinely with SRT data. Also, this would be desirable to record biological variables such as sex, race, and age while capturing gene expression in parallel. We envisage that accounting for such variations across individuals in SRT data and existing histology images will revolutionize the future of disease identification.\
		
		In analyzing SRT data, the pre-processing step can dramatically affect the results. The captured location undergoes sequencing for generating SRT data to generate the count matrix known as the gene expression matrix. However, barely acknowledged explicitly, the obtained count data from sequencing machines has the compositional nature for which the abundance of each gene can be described as proportions or probabilities to other genes within that sample \cite{compositional}. Subsequently, the gene expression matrix in SRT data as a compositional data exists in a non-Euclidean space, but rather in a sub-space known as the simplex \cite{aitchison1982statistical}. In the simplex space, a proposed alternative is the Aitchison distance. Yet, using transformation methods such as log-ratio transformation, the compositional data is mapped into real space \cite{aitchison1982statistical}, making the Euclidean distance meaningful. Thus, applying conventional analyses, including dimensional reduction and statistical methods, without appropriate transformation and normalization strategies can lead to misinterpretation of the data (refer to \cite{compositional} for more information about proving compositionally in sequencing data). Although the reviewed techniques in this study mainly leverage the Euclidean distance for the spatial coordinates, which is completely meaningful, using the PCA or clustering algorithms (i.e., K-means) on the untransformed data is against the compositional data hypothesis. Among the reviewed techniques, 13 methods have considered log-transformation on gene expression matrix, and the remaining performed only the normalization method on the gene expression matrix. For example, in the stLearn approach, the authors proposed the SMEClust normalization, which performs PCA and UMAP on the normalized genes without any transformation. Nonetheless, we would suggest that the future works should take account of compositionality in gene expression matrix and investigate the other transformation and normalization approaches.\\
		Regardless of the technology that provides the SRT data, the gene expression matrix is in sparse form, consisting of an excessive amount of zero values, displaying appreciable overdispersion, which makes it challenging to find an appropriate model for the count data. Since many statistical and DL models (i.e., VAE models) directly model the count data, understanding the overdispersion and zero inflation patterns of gene expression is essential. Also, the fitted model can give an overall view of whether this sparsity is caused by the  platform (which there is a demand for an imputation method) or whether it is caused by the gene expression heterogeneity across tissue locations (which need the model to account for overdispersion and zero inflation) \cite{Zhao2022}. Zhao et al. \cite{Zhao2022} showed that the preferable models across most SRT technologies are the Poisson or the negative binomial (NB) models with direct modeling of overdispersion without an additional zero inflation term. Additionally, they expressed that the excessive zero count in SRT data potentially reflects biological variation, which imputation methods can highly generate noises by adding non-zero values, negatively impacting the analysis. Therefore, we strongly suggest further assessment of the current existing imputation methods such as gimVI (also the ZINB model is used to model the count data) and Tangram to consider the limitations mentioned above.\\
		In addition, All methods mentioned, are reference-based, meaning they rely on an external scRNA-seq dataset from the same tissue to estimate cell proportions. Chen et al. conducted a comprehensive analysis of computational methods for cell-type deconvolution using internal inference (using the single-cell resolution SRT dataset) and external inference (using the scRNA-seq dataset from the same tissue as reference) while also investigating the impact of gene-subset selection. The results showed that Tangram and DSTG performed best with perfectly matched internal references and that the performance of deconvolution can be affected by the selection of genes. Most methods performed better with top cell-type marker genes compared to highly variable gene (HVG) subsets in the case of external reference. In analyzing single-cell resolution transcriptomics datasets, dimension reduction techniques such as PCA are a crucial step, followed by clustering algorithms, which consider different loss functions. The reviewed techniques in this study treated dimension reduction methods as error-free techniques for obtaining low-dimensional features. It would be valuable for future studies to develop a new method that combines dimension reduction and clustering with a unified loss function, and to evaluate the performance of dimension reduction approaches. \cite{DR-SC}.\\
		Assay for Transposase Accessible Chromatin with high-throughput sequencing (ATAC-seq) is a technology designed to identify genome-wide profiling of chromatin accessibility \cite{baek2020single}. Regarding the emergence of single-cell biology, single-cell ATAC sequencing (scATAC-seq) has provided chromatin accessibility at single-cell resolution.
		The current epigenomic profiling methods lack spatial resolution; several methods \cite{llorens2023solid,deng2022spatial} have been developed to perform spatially resolved chromatin accessibility profiling (spatial ATAC) in animal and human tissue sections. However, obtaining spatial information for these technologies mainly need a higher resolution. Recently, VAEs have been utilized to learn joint latent space for gene imputation tasks such as gimVI and understand the phenotypic interplay between gene expression and TCR sequence \cite{an2021jointly}. Moreover, the integrating and providing a unified view of multi-omics has received considerable attention for researchers to jointly profile the transcriptional and chromatin land-scape of single-cells, such as \cite{cao2022multi,ashuach2021multivi,Lotfollahi2022.03.16.484643}. Thus, developing the DL models for integrating scATAC-seq and spatial ATAC-seq to learn the latent embedding jointly would be innovative \cite{park2022spatial}. 
		
		\section{Conclusion}
		In conclusion, several innovative deep learning (DL) approaches have been developed to address single-cell resolution transcriptomics (SRT) data analysis challenges. DL algorithms are well-suited to these challenges because they can uncover complex patterns and efficiently analyze large and multi-modal data. As SRT data grows and diversifies, software and DL approaches that can effectively tackle and interrogate multimodal SRT data will be in high demand. In this paper, we thoroughly reviewed all DL methods and the challenges they address in analyzing spatially resolved transcriptomics data. We categorized these methods into six main categories based on their main task and downstream analysis, including identifying the spatial domain, identifying spatially variable genes, imputing missing genes, improving gene expression resolution, analyzing cell-cell interactions, and performing cell-type decomposition. We hope this review serves as a comprehensive reference for guiding the usage of  DL methods in SRT data analysis and encourages scholars with complementary expertise to collaborate and develop new methods that integrate gene expression, spatial information, single-cell data, and digital pathology to drive innovation in these spatial technologies.
		\onecolumn
		\begin{landscape}
			\begingroup
			\footnotesize
			\begin{longtable}{c l l p{4cm} p{4cm} c p{3.5cm}}
				\caption{Deep Learning Algorithms.}\\
				\label{papers}
				\textbf{Algorithm} &\textbf{Category}&\textbf{Models}& \textbf{Advantages}& \textbf{Limitations}
				& \textbf{Data Type}&\textbf{Codes}\\
				
				\endhead
				\hline
				gimVI\cite{lopez2019joint}&Imputing missing genes&VAE& Can jointly use scRNA-seq and spatial transcriptomics to impute missing genes in SRT data.& Evaluate the model on the small part of genes, which obtain relatively low Spearman correlation.&\Shortunderstack{osmFISH\\starMAP}&\Shortunderstack{https://github.com\\/YosefLab/scVI}\\
				\hline
				CoSTA\cite{xu2021costa}&Identifying SVG&
				CNN&
				Learn broader spatial patterns
				&High parameters tuning& In Situ Hybridization&\Shortunderstack{
					https://github.com\\/rpmccordlab/CoSTA}
				\\
				\hline
				SpaCell\cite{tan2019spacell}&Identifying spatial domain&
				\Shortunderstack{CNN\\AE\\DNN}&
				Can incorporate the three types of spatial 
				transcriptomics data:histology,imaging, and gene expression
				&Pre-trained on the ImageNet which is unrelated to SRT &Slide-seq
				&\Shortunderstack{https://github.com/\\BiomedicalMachine\\Learning/SpaCell}
				\\
				\hline
				stlearn\cite{pham2020stlearn}&\Shortunderstack{Identifying spatial domain\\cell-cell interactions}&
				CNN&
				Combine three various type of data
				&Lack of quantitative comparison&10x Genomics&\Shortunderstack{
					https://github.com/\\petersaj/histology}\\
				\hline
				SpaGCN\cite{hu2020integrating}&\Shortunderstack{Identifying spatial domain\\Identifying SVG}&GCN&Can incorporate the three types of the spatial transcriptomics data:histology,imaging, and gene expression. can detect SVGs. &Using the RGB channel to analyze the histology images may not be appropriate in noisy images. Inadequate reasons to show the influence of the histology images in ST methods. &\Shortunderstack{smFISH\\Slide-Seq}&\Shortunderstack{https://github.com/\\
					jianhuupenn/SpaGCN}\\
				\hline
				SEDR\cite{fu2021unsupervised}&Identifying spatial domain&\Shortunderstack{GCN\\AE\\DNN\\VGAE}&Can learn the low-dimension representation of gene expression jointly with embedding spatial information & Defining the similarity between spots before training and did not consider the learning strategy &\Shortunderstack{10x Genomics\\Stereo-seq}&\Shortunderstack{https://github.com\\/HzFu/SEDR}\\
				\hline
				STAGATE\cite{dong2021deciphering}&\Shortunderstack{Identifying spatial domain\\Identifying SVG}&GAT&Can adaptively learn the similarity between spots and construct SNN based on the adjacency matrix and cell-type aware module. First method for constructing a 3D pattern with spatial information.&Rely on a radius parameter
				to determine the adjacency matrix &\Shortunderstack{10x
					Visium\\Slide-seq\\ Slide-seqV2\\Stereo-seq} &\Shortunderstack{http://spatial.libd.org\\/spatialLIBD}\\
				\hline
				RESEPT\cite{chang2021define}&Identifying spatial domain&\Shortunderstack{GCN\\CNN\\AE}&Can reconstruct RGB image from gene expression or RNA velocity from SRT data.& Use the fixed number of neighbors to build the input graph.&10x Visium&\Shortunderstack{ https://github.com/\\OSUBMBL/RESEPT}\\
				\hline
				ST-Net\cite{he2020integrating}&Identifying SVG&CNN&Can link gene expression with visual features. Can generalize in the other breast cancer spatial transcriptomics data due to the promising results in external validation.& Spatial information is not utilized in the model.& In situ sequencing& \Shortunderstack{https://github.com/\\bryanhe/ST-Net}\\ 
				\hline
				HisToGene\cite{pang2021leveraging}&\Shortunderstack{Identifying SVG\\Enhancement of GER}& MHA& Can join histology images with spatial information to predict gene expression;first paper for high-resolution gene expression prediction.& Poor performance in low-resolution SRT data.&10x Visium&\Shortunderstack{https://github.com/\\maxpmx/HisToGene}\\
				\hline
				SPADE\cite{bae2021discovery}&Identifying SVG&CNN&Can identify maker genes associated with morphology. Can use gene ontology to find the relationship between obtained principal components from image features and biological processes.&The obtained marker genes can be significantly affected by the spot's density and the distance between spots.& 10x Visium&\Shortunderstack{https://github.com/\\mexchy1000/
					spade}\\
				\hline
				ECNN\cite{chelebian2021morphological}&Identifying spatial domain&CNN&Can pre-train CNN on the relative dataset instead of using ImageNet.& The proposed model is blind to SRT data&10x Visium&-\\
				\hline
				Tangram\cite{biancalani2021deep}&Imputing missing genes&CNN& Can provide an automatic pipeline for locating histology data on an anatomically annotated CCF.& It requires a CCF, which is available for a few organs related to the mouse brain.&In situ hybridization&\Shortunderstack{https://github.com/\\broadinstitute/Tangram}\\
				\hline
				GIST\cite{zubair2021jointly}&Cell-type decomposition&CNN&Can use the image as informative prior information to improve cell-type decomposition&Did not consider the spatial data. It Needs to tune the hyper-parameter $\lambda$ empirically.&In Situ Sequencing&\Shortunderstack{https://github.com/\\asifzubair/GIST}\\
				\hline
				CNNTL\cite{pegah}&Identifying SVG&CNN& Can evaluate the CNN model on the ISH images to predict gene expression and utilize triplet loss to overcome the lack of enough labels for each gene.& The method is limited to a small fraction of genes in brain layers. The model pre-trained on the datasets unrelated to SRT data.& In situ hybridization& \Shortunderstack{https://github.com/PegahA/\\Human\_Brain\_ISH\_
					ML}\\
				\hline
				XFuse\cite{xfuse}&Enhancement of GER&VAE&Can find a clear pattern of low-resolution SRT data and impute missing genes at high-resolution.& Only detect genes whose spatial patterns are similar to the histology images.&In situ RNA capturing&\Shortunderstack{https://github.com/\\ludvb/xfuse}\\
				\hline
				JSTA\cite{littman2021joint}&Identifying spatial domain&DNN&Can enhance the cell segmentation at the cell borders by jointly using cell-type expression patterns and RNA hybridization-based spatial transcriptomics.& The method was not compared with the other ML methods. It uses parameter tuning empirically to identify cell borders.&\Shortunderstack{MERFISH\\osmFISH}&\Shortunderstack{https://github.com/\\wollmanlab/JSTA}\\
				\hline
				DSTG\cite{DSTG}&Cell-type decomposition&GCN& Apply semi-supervised graph convolution network to detect cell type deconvolution in ST data. It generates pseudo-ST data by scRNAseq data and provides ground truth for the learning process.& Considering the Euclidean distance between the pseudo-ST and real-ST data to show the similarity between two data is not a fair comparison.& \Shortunderstack{10X Genomics
					Visium\\Slide-seq v2}&\Shortunderstack{https://github.com/\\Su-informatics-lab/DSTG}\\
				\hline
				DeepSpaCE\cite{deepspace}&\Shortunderstack{Identifying SVG\\Enhancement of GER}&CNN&Use Semi-supervised learning to enhance the CNN model. Perform super-resolution and section imputation methods, which reduce the experimental costs.&It predicts only limited genes. The model only can be applied to the images.&In situ capturing&\Shortunderstack{https://github.com/\\tmonjo/DeepSpaCE}\\
				\hline
				GCNG\cite{yuan2020gcng}&Cell-Cell interactions&GCN&Can identify extracellular interaction in SRT data, even in cells without direct relationship.& The model relies on the pre-defined distance criteria for selecting the neighbor cells.&\Shortunderstack{seqFISH+\\MERFISH}&\Shortunderstack{https://github.com\\/xiaoyeye/GCNG}\\
				\hline
				conST\cite{zong2022const}&\Shortunderstack{Identifying spatial domain\\Identifying SVG\\Cell-Cell interactions}&\Shortunderstack{GCN\\VGAE\\AE}&It uses multi-modal contrastive learning that can be utilized in both SRT categories.&High parameters tuning.&\Shortunderstack{seqFISH\\MERFISH\\10x Visium\\Stereo-seq}&\Shortunderstack{https://github.com\\/ys-zong/conST}\\
				\hline
			\end{longtable}
			\endgroup
		\end{landscape}
		
		\twocolumn
		
		\textcolor{red}{\section{Key Points}
			\begin{enumerate}
				\item Spatially resolved transcriptomics (SRT) is a new technology providing the position of captured expression across the tissue at single-cell level resolution.
				\item Twenty-one deep learning-based methods are systemically reviewed in this paper and categorized into six main groups based on the tasks and downstream analyses.
				\item A brief discussion of the current machine learning approaches are presented for each category to assess the advantages of deep learning models proposed for that category in comparison to the traditional machine learning models.
				\item A unified description of the model and result corresponding to each deep learning  models is presented, and the mathematical model is also discussed in the supplementary section.
				\item Lastly, a comprehensive summary of the deep learning algorithm, evaluation metrics, and datasets by each approach is tabulated.
		\end{enumerate}}
		\section{Competing interests}
		No competing interest is declared.
		
		\section{Author contributions statement}
		
		HAR designed the study; RZN wrote the manuscript; HAR, RZN, RE, CM, AB, AA, NL, and ML edited the manuscript; RZN generated all figures and tables. HAR provided his advice in generating figures and tables. All authors have read and approved of the final version of the paper.
		\section{FUNDING}
		This work was supported by a UNSW Scientia Program Fellowship; and the Australian Research Council Discovery Early Career Researcher Award (DECRA) [DE220101210 to HAR].
		\section{Acknowledgments}
		We thank Associate Prof Nadeem Kaakoush (School of Biotechnology and Biomolecular Sciences, UNSW Sydney), for their constructive comments on the manuscript.



		
		
	\end{document}